\documentclass[a4paper,11pt]{article}
\usepackage{jheppub} 
\usepackage[T1]{fontenc}
\usepackage{booktabs}
\usepackage[dvips,table]{xcolor} 
\usepackage{xspace}
\usepackage{dcolumn}
\usepackage{caption}
\usepackage
[subrefformat=parens,position=top,skip=-15pt,margin=15pt,justification=justified,singlelinecheck=false]
{subcaption}

\newcommand{\Pl}{\ell}

\def\reffi#1{\mbox{Figure~\ref{#1}}}
\def\reffis#1{\mbox{Figures~\ref{#1}}}
\def\refta#1{\mbox{Table~\ref{#1}}}

\def\refse#1{\mbox{Section~\ref{#1}}}
\def\citere#1{\mbox{Ref.~\cite{#1}}}
\def\citeres#1{\mbox{Refs.~\cite{#1}}}

\newcommand{\ri}{\mathrm i}

\def\be{\begin{equation}}
\def\ee{\end{equation}}

\newcommand{\PH}{\ensuremath{\text{H}}\xspace}
\newcommand{\Pj}{\ensuremath{\text{j}}\xspace}
\newcommand{\Pp}{\ensuremath{\text{p}}\xspace}
\newcommand{\Pe}{\ensuremath{\text{e}}\xspace}
\newcommand{\Pb}{\ensuremath{\text{b}}\xspace}

\newcommand{\Pt}{\ensuremath{\text{t}}\xspace}
\newcommand{\Pu}{\ensuremath{\text{u}}\xspace}
\newcommand{\Pd}{\ensuremath{\text{d}}\xspace}
\newcommand{\Ps}{\ensuremath{\text{s}}\xspace}
\newcommand{\Pc}{\ensuremath{\text{c}}\xspace}
\newcommand{\Pg}{\ensuremath{\text{g}}\xspace}

\newcommand{\PW}{\ensuremath{\text{W}}\xspace}
\newcommand{\PZ}{\ensuremath{\text{Z}}\xspace}

\newcommand{\Mt}{\ensuremath{m_\Pt}\xspace}
\newcommand{\MH}{\ensuremath{M_\PH}\xspace}
\newcommand{\MWOS}{\ensuremath{M_\PW^\text{OS}}\xspace}
\newcommand{\MW}{\ensuremath{M_\PW}\xspace}
\newcommand{\MZOS}{\ensuremath{M_\PZ^\text{OS}}\xspace}
\newcommand{\MZ}{\ensuremath{M_\PZ}\xspace}

\newcommand{\Gt}{\ensuremath{\Gamma_\Pt}\xspace}

\newcommand{\GZOS}{\ensuremath{\Gamma_\PZ^\text{OS}}\xspace}
\newcommand{\GW}{\ensuremath{\Gamma_\PW}\xspace}
\newcommand{\GWOS}{\ensuremath{\Gamma_\PW^\text{OS}}\xspace}

\newcommand{\GeV}{\ensuremath{\,\text{GeV}}\xspace}
\newcommand{\TeV}{\ensuremath{\,\text{TeV}}\xspace}

\newcommand{\alphas}{\ensuremath{\alpha_\text{s}}\xspace}
\newcommand{\order}[1]{\ensuremath{\mathcal{O}{\left(#1\right)}}\xspace}

\newcommand{\abs}[1]{\left|#1\right|}

\newcommand{\GF}{\ensuremath{G_\mu}}

\newcommand{\pt}{\ensuremath{p_\text{T}}\xspace}
\newcommand{\ptsub}[1]{\ensuremath{p_{\text{T},#1}}\xspace}

\renewcommand{\Re}{\mathop{\mathrm{Re}}\nolimits}

\newcommand{\fullProcess  }{\ensuremath{\Pp\Pp\to\Pe^+\nu_\Pe \mu^-\bar{\nu}_\mu\Pb\bar{\Pb}\PH}\xspace}
\newcommand{\ggProcess    }{\ensuremath{\Pg\Pg\to\Pe^+\nu_\Pe \mu^-\bar{\nu}_\mu\Pb\bar{\Pb}\PH}\xspace}
\newcommand{\qqbProcess    }{\ensuremath{q\bar{q}\to\Pe^+\nu_\Pe \mu^-\bar{\nu}_\mu\Pb\bar{\Pb}\PH}\xspace}

\newcommand{\recola}{{\sc Recola}\xspace}
\newcommand{\collier}{{\sc Collier}\xspace}
\newcommand{\madgraph}{{\sc\small MadGraph5\_aMC@NLO}\xspace}

\newcolumntype{.}{D{.}{.}{-1}}
\newcolumntype{d}[1]{D{.}{.}{#1}}

\renewcommand{\vec}[1]{\mathbf{#1}}
\colorlet{tableoverheadcolor}{gray!37.5}
\colorlet{tableheadcolor}{gray!25}
\colorlet{tablerowcolor}{gray!12.5}

\newlength{\width}
\newlength{\height}
\newcommand{\brabar}[1]{%
    \settoheight{\height}{\ensuremath{#1}}%
    \settowidth{\width}{\ensuremath{#1}}%
    \makebox[0pt][l]{\ensuremath{#1}}%
    \raisebox{1.26ex}{\scalebox{.3}{\textbf{(}}}%
    \rule[1.41\height]{0.7\width}{0.35pt}%
    \raisebox{1.26ex}{\scalebox{.3}{\textbf{)}}}%
}

\newcommand{\change}[1]{{#1}}

\title{NLO QCD corrections to off-shell top--antitop production with leptonic decays in association with a Higgs boson at the LHC}
\subheader{\today}

\author{Ansgar Denner,}
\author{Robert Feger}

\affiliation{%
        Universit\"at W\"urzburg, %
        Institut f\"ur Theoretische Physik und Astrophysik, %
        Emil-Hilb-Weg 22, \linebreak %
        97074 W\"urzburg, %
        Germany%
}
\emailAdd{ansgar.denner@physik.uni-wuerzburg.de}
\emailAdd{robert.feger@physik.uni-wuerzburg.de}

\abstract{%
  We compute the hadronic production of top--antitop pairs in
  association with a Higgs boson at next-to-leading-order QCD,
  including the decay of the top and antitop quark into bottom quarks
  and leptons. Our computation is based on full leading and
  next-to-leading-order matrix elements for $\Pe^+\nu_\Pe
  \mu^-\bar{\nu}_\mu\Pb\bar{\Pb}\PH(\Pj)$ and includes all
  non-resonant contributions, off-shell effects and interferences.
  Numerical results for the integrated cross section and several
  differential distributions are given for the LHC operating at 13\TeV
  using a fixed and a dynamical factorization and renormalization
  scale. The use of the dynamical instead of the fixed scale improves
  the perturbative stability in high-energy tails of most
  distributions, while the integrated cross section is hardly affected
  differing by only about one per cent and leading to \change{almost the same
  $K$~factor of about 1.17.}}

\begin{document}

\maketitle
\flushbottom

\section{Introduction}
\label{sec:introduction}

After the discovery of a new boson of a mass around 125\GeV by the CMS
and ATLAS Collaborations \cite{Aad:2012tfa,Chatrchyan:2012ufa} at the
LHC, the determination of its quantum numbers and couplings to other
particles has become a high priority in particle physics. Results from
the first run of the LHC strongly support the hypothesis that this
particle is the Higgs boson predicted by the Standard Model (SM) of
particle physics. In the Brout--Englert--Higgs symmetry breaking
mechanism of the SM the Higgs boson is the key for understanding the
origin of mass. Since in this framework the Higgs boson couples to
fermions with a strength proportional to their mass via Yukawa
interactions, its coupling to the heaviest quark, the top quark, is of
particular interest. The main production mechanism of the Higgs boson
in the SM is gluon fusion, $\Pg\Pg{\to}\PH$. This process is
sensitive to the top-quark Yukawa coupling, but possible heavy
particles beyond the SM running in the loop bias its determination.
On the other hand, the production of a SM Higgs boson in
association with a top-quark pair allows a direct access of the
top-quark Yukawa coupling already at tree level, disentangling it from
possible beyond-SM contributions.

Leading-order (LO) predictions for $\Pt\bar\Pt\PH$ production for
stable Higgs boson and top quarks have been presented in
\citeres{Raitio:1978pt,Ng:1983jm,Kunszt:1984ri,Gunion:1991kg,Marciano:1991qq}.
Since LO predictions of QCD processes suffer from large perturbative
uncertainties higher-order corrections have to be taken into account
for adequate theoretical predictions. The cross section for
$\Pt\bar\Pt\PH$ production at next-to-leading-order (NLO) QCD is known
for more than 10 years
\cite{Beenakker:2001rj,Beenakker:2002nc,Reina:2001sf,Dawson:2002tg,
  Dawson:2003zu}. Meanwhile, NLO QCD corrections have been matched to
parton showers \change{\cite{Frederix:2011zi,Garzelli:2011vp,Hartanto:2015uka}} and recently
electroweak corrections to $\Pt\bar\Pt\PH$ production have been
computed \cite{Frixione:2014qaa,Yu:2014cka,Frixione:2015zaa}. 
\change{NLO QCD corrections for the 
important background processes $\Pt\bar\Pt\Pb\bar\Pb$ and  $\Pt\bar\Pt\Pj\Pj$
production have been worked out in 
\citeres{Bredenstein:2008zb,Bredenstein:2009aj,Bevilacqua:2009zn,Bredenstein:2010rs}
and \citeres{Bevilacqua:2010ve,Bevilacqua:2011aa,Bevilacqua:2014qfa}, respectively, 
and matched to parton showers in
\citeres{Kardos:2013vxa,Cascioli:2013era,Garzelli:2014aba} and 
\citere{Hoeche:2014qda}.}

Several
searches for the associated production of the Higgs boson with a
top-quark pair for a variety of decay channels have been published by
ATLAS
\cite{ATLAS:2012cpa,ATLAS:2014,Aad:2014lma,ATLAS-CONF-2015-006,Aad:2015gra}
and CMS
\cite{CMS:2012qaa,CMS:2013sea,CMS-PAS-HIG-13-015,Chatrchyan:2013yea,
  Khachatryan:2014qaa,Khachatryan:2015ila}. These searches are
challenging due to a large background from $\Pt\bar\Pt\Pb\bar\Pb$ and
$\Pt\bar\Pt\Pj\Pj$ production, and so far no evidence of a
$\Pt\bar\Pt\PH$ signal over background has been found. The current
ratio of the measured $\Pt\bar\Pt\PH$ signal cross section to the SM
expectation quoted by ATLAS is $\mu=1.5\pm 1.1$ \cite{Aad:2015gra} and
by CMS $\mu=1.2^{+1.6}_{-1.5}$ \cite{Khachatryan:2015ila} for a
Higgs-boson mass of 125\GeV.

In this article we present the calculation of the NLO QCD corrections
to the hadronic production of a positron, a muon, missing energy, two
\Pb~jets and a SM Higgs boson, which we assume to be stable
($\fullProcess$) at the 13\TeV LHC, which includes the resonant
production of a top--antitop-quark pair in association with a Higgs boson with
a subsequent leptonic decay of the top and the antitop quark. 
\change{Our calculation includes all NLO QCD correction effects in
  $\Pt\bar\Pt\PH$ production and top decays and} 
also
takes into account all off-shell, non-resonant and interference
effects of the top quarks. We consider the fixed renormalization and
factorization scale used
in \citere{Beenakker:2002nc} and alternatively the dynamical scale
choice from \citere{Frederix:2011zi} and investigate their quality in
reducing the dependence of the integrated cross section as well as
differential distributions on the factorization and renormalization
scales. The phase-space integration is performed with a newly
implemented in-house multi-channel Monte Carlo program, using
phase-space mappings similar to \citere{Dittmaier:2002ap}. The Monte
Carlo implements the dipole subtraction method
\cite{Catani:1996vz,Dittmaier:1999mb,Catani:2002hc,Phaf:2001gc} for the computation of the real
corrections and is linked to the matrix-element
generator \recola \cite{Actis:2012qn} for the computation of the
LO and NLO matrix elements as well as colour- and
spin-correlated squared matrix elements needed for the evaluation of
subtraction terms.

The calculation follows in many respects the one of
$\Pp\Pp\to\Pe^+\nu_\Pe \mu^-\bar{\nu}_\mu\Pb\bar{\Pb}$ in
\citere{Denner:2012yc}. In particular, since the additional Higgs
boson in the final state has no colour charge, the contributing
Catani--Seymour dipoles \cite{Catani:1996vz,Catani:2002hc} are the
same. For the hadronic process $\Pp\Pp\to\Pe^+\nu_\Pe
\mu^-\bar{\nu}_\mu\Pb\bar{\Pb}\PH$ we find a qualitative similar
behaviour of NLO corrections to the integrated cross section as well
as for differential distributions of the same observables. Where
appropriate we compare results and point out differences. Moreover, we
compare our results for the total cross sections to those of existing
calculations for on-shell Higgs boson and top quarks
\cite{Beenakker:2002nc,Frederix:2011zi}.

The paper is organised as follows. In \refse{sec:calculation} we discuss details 
of the calculation such as contributing subprocesses and Feynman diagrams and 
explain some technical aspects of the real (\refse{ssec:RealCorrections}) and 
virtual (\refse{ssec:VirtualCorrections}) corrections. We present numerical 
results for the LHC operating at $\sqrt{s}=13\TeV$ in \refse{sec:results}. In 
\refse{ssec:InputParameters} we list and explain the input parameters, jet 
definition, cuts and scale choices for our calculation, while results for 
integrated cross sections and a discussion of the scale dependence are provided 
in \refse{ssec:IntegratedCrossSectionScaleDependence}. In 
\refse{ssec:DifferentialDistributions} we display and discuss several 
differential distributions for a fixed and a dynamical scale choice. We have 
performed several checks of our calculation which we present in 
\refse{sec:Checks}. Our conclusions are given in \refse{sec:Conclusions}.

\section{Details of the calculation}
\label{sec:calculation}

\begin{figure}
        \newcommand{\myframebox}{\framebox}
        \renewcommand{\myframebox}{\relax}
        \setlength{\parskip}{-10pt}
        \captionsetup[subfigure]{margin=0pt}
        \begin{subfigure}{0.32\linewidth}
                \subcaption{}
                \myframebox{
                        \includegraphics[width=\linewidth]{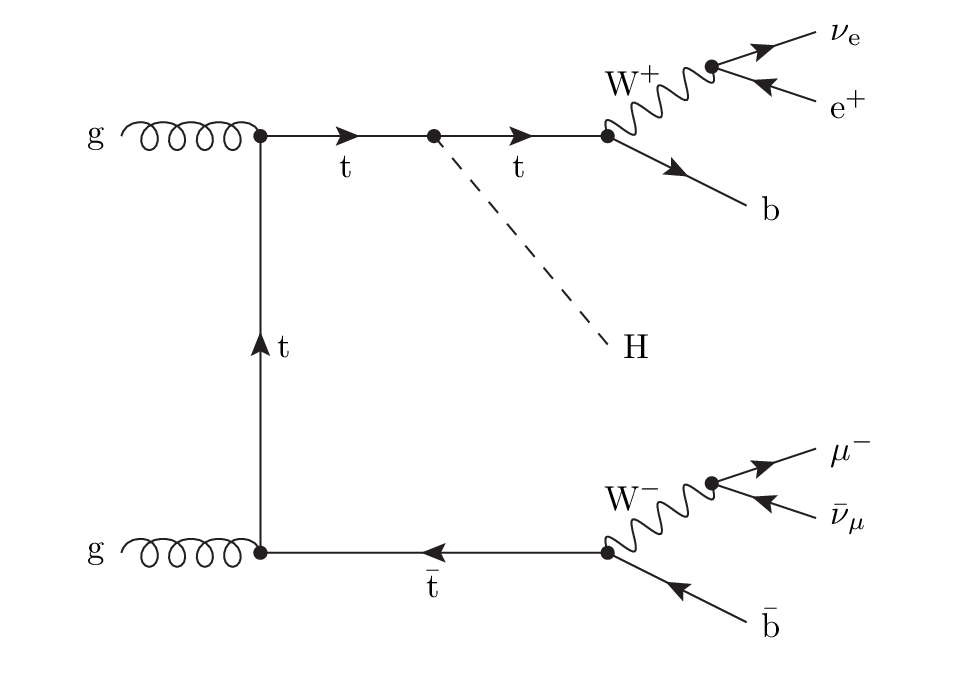}
                }
                \label{fig:born_2tops_gg_tchannel} 
        \end{subfigure}
        \begin{subfigure}{0.32\linewidth}
                \subcaption{}
                \myframebox{
                        \includegraphics[width=\linewidth]{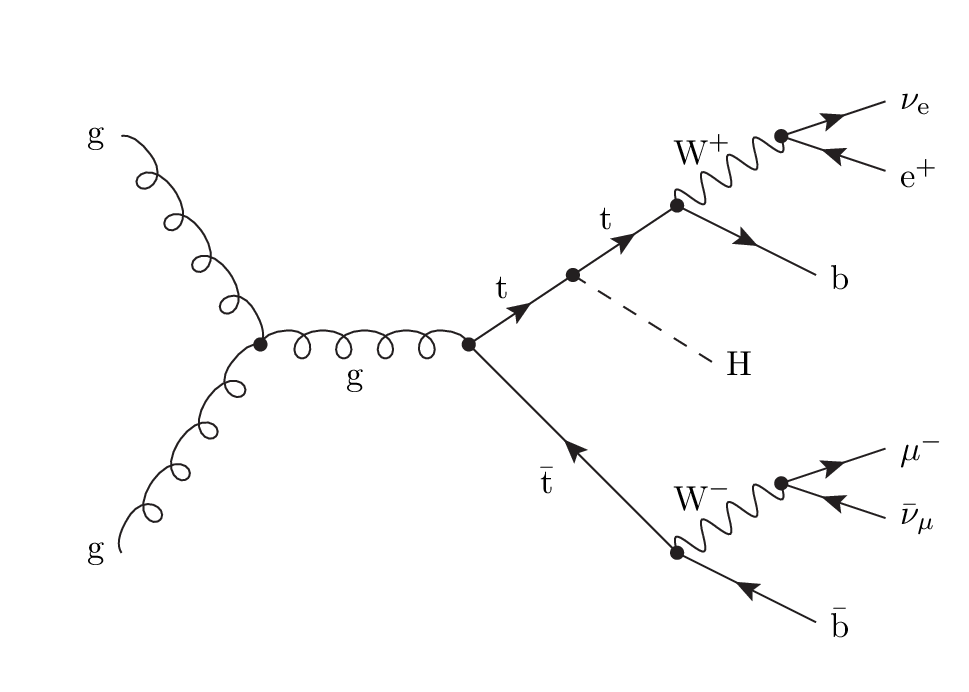}
                }
                \label{fig:born_2tops_gg_schannel} 
        \end{subfigure}
        \begin{subfigure}{0.32\linewidth}
                \subcaption{}
                \myframebox{
                        \includegraphics[width=\linewidth]{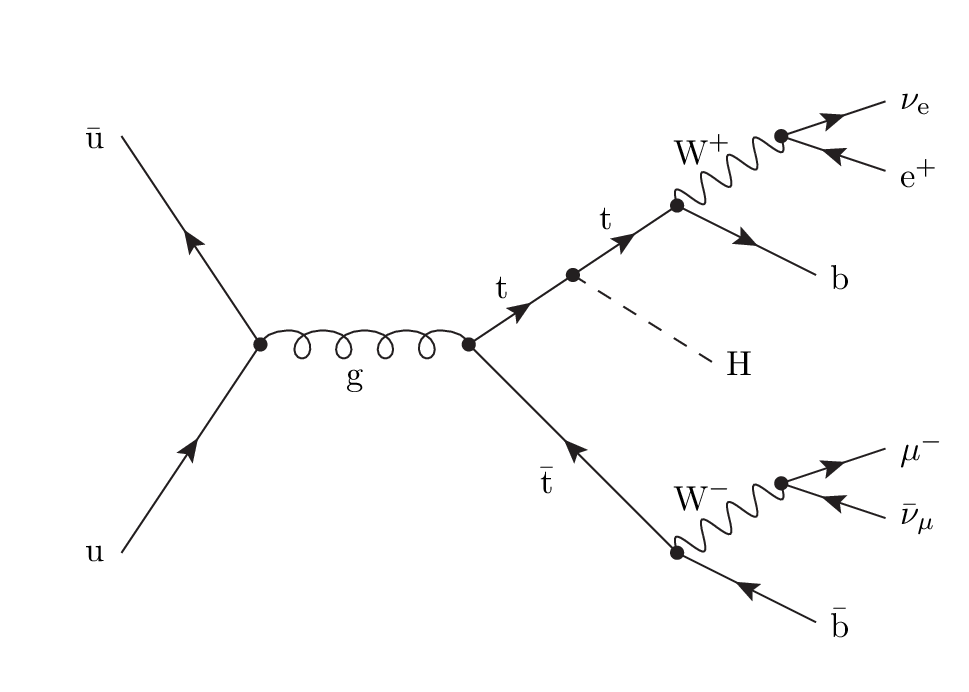}
                }
                \label{fig:born_2tops_uxu_schannel} 
        \end{subfigure}

        \begin{subfigure}{0.32\linewidth}
                \subcaption{}
                \myframebox{
                        \includegraphics[width=\linewidth]{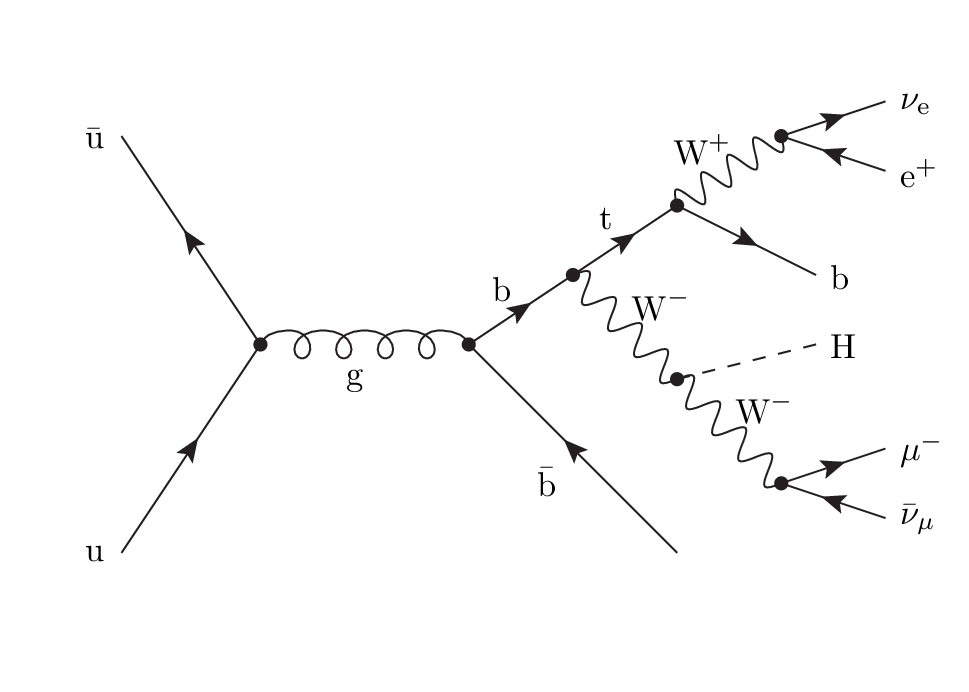}
                }
                \label{fig:born_1top_uxu_schannel} 
        \end{subfigure}
        \begin{subfigure}{0.32\linewidth}
                \subcaption{}
                \myframebox{
                        \includegraphics[width=\linewidth]{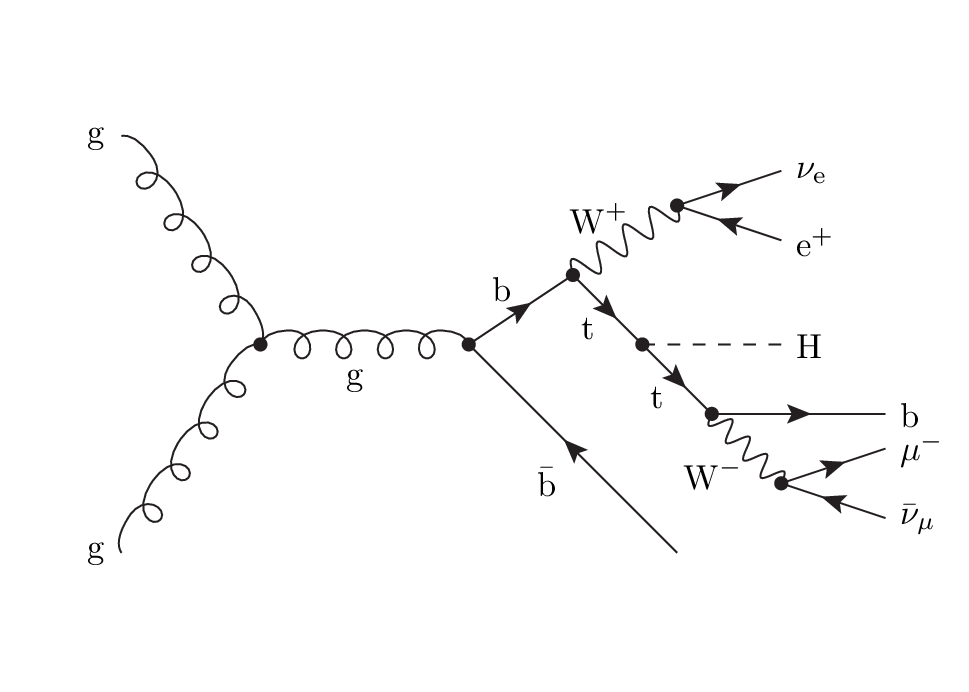}
                }
                \label{fig:born_1top_gg_schannel} 
        \end{subfigure}
        \begin{subfigure}{0.32\linewidth}
                \subcaption{}
                \myframebox{
                        \includegraphics[width=\linewidth]{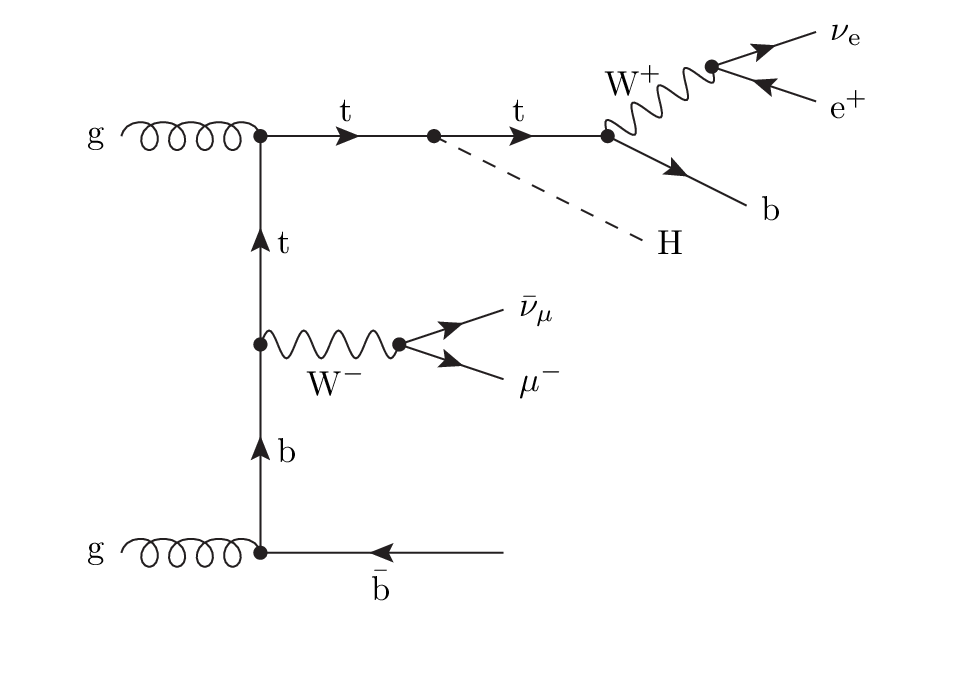}
                }
                \label{fig:born_1top_gg_tchannel} 
        \end{subfigure}
        
        \begin{subfigure}{0.32\linewidth}
                \subcaption{}
                \myframebox{
                        \includegraphics[width=\linewidth]{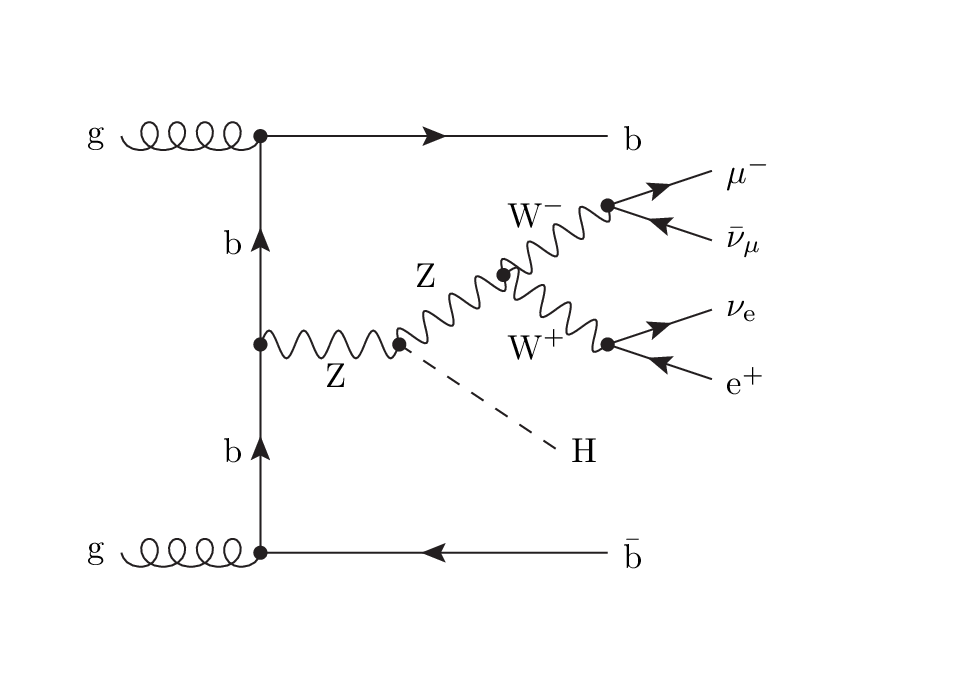}
                }
                \label{fig:born_notops_gg_tchannel_z0} 
        \end{subfigure}
        \begin{subfigure}{0.32\linewidth}
                \subcaption{}
                \myframebox{
                        \includegraphics[width=\linewidth]{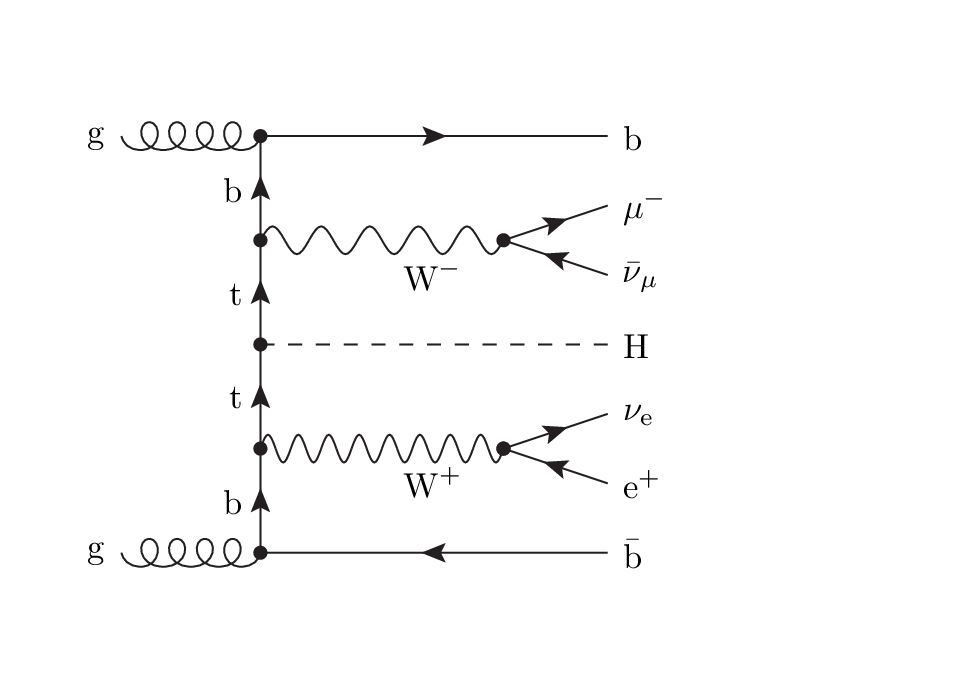}
                }
                \label{fig:born_notops_gg_tchannel} 
        \end{subfigure}
        \begin{subfigure}{0.32\linewidth}
                \subcaption{}
                \myframebox{
                        \includegraphics[width=\linewidth]{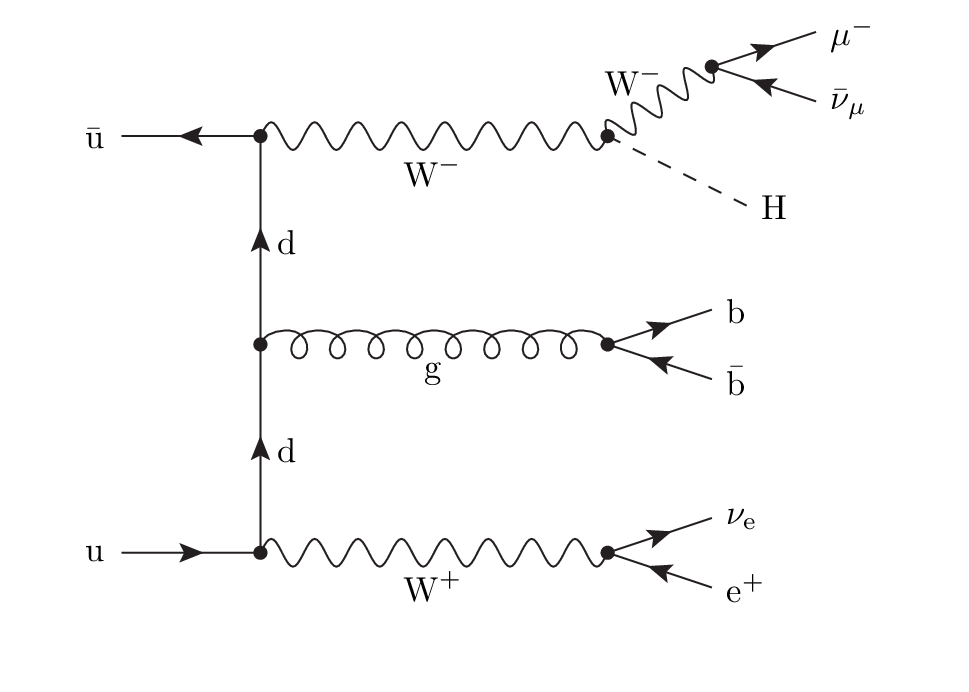}
                }
                \label{fig:born_notops_uxu_tchannel} 
        \end{subfigure}

        \caption{\label{fig:lo_tree_feynman_diagrams}%
                Representative tree-level Feynman diagrams with %
                \subref{fig:born_2tops_gg_tchannel}--\subref{fig:born_2tops_uxu_schannel} two (top), %
                \subref{fig:born_1top_uxu_schannel}--\subref{fig:born_1top_gg_tchannel} one (middle) and %
                \subref{fig:born_notops_gg_tchannel_z0}--\subref{fig:born_notops_uxu_tchannel} no top-quark resonances (bottom).%
        }
\end{figure}

We compute the QCD corrections to the full hadronic process
\begin{equation}\label{eqn:full_process}
        \fullProcess.
\end{equation}
We consider the tree-level amplitude 
at $\order{\alphas\alpha^{5/2}}$ including all 
resonant, non-resonant, and off-shell effects of the top quarks and
all interferences. 
Neglecting flavour mixing as well as contributions from the suppressed 
bottom-quark parton densities and counting \Pu, \Pd, \Pc and \Ps quarks 
separately, we distinguish 5 partonic channels for the LO hadronic 
process: the gluon-induced process \ggProcess and four processes from 
\qqbProcess by substituting different quark flavours ($q=\Pu, \Pd, \Pc, \Ps$). 
Throughout this paper we consider the bottom quark massless, implying no 
contribution from tree diagrams involving the Higgs--bottom-quark coupling. The 
$\Pg\Pg$ process involves 236 and the $q \bar{q}$ processes 98 tree diagrams 
each under these prerequisites. In \reffi{fig:lo_tree_feynman_diagrams} we show 
sample diagrams grouped by the number of top-quark resonances.

\subsection{Real corrections}
\label{ssec:RealCorrections}

The real correction process $\Pp\Pp\to\Pe^+\nu_\Pe 
\mu^-\bar{\nu}_\mu\Pb\bar{\Pb}\PH\Pj$ receives contributions from the 13 partonic 
subprocesses
\begin{equation}
        \begin{aligned}
                \Pg\Pg      & \to \Pe^+\nu_\Pe \mu^-\bar{\nu}_\mu\Pb\bar{\Pb}\PH\Pg,\\
                q \bar{q}   & \to \Pe^+\nu_\Pe \mu^-\bar{\nu}_\mu\Pb\bar{\Pb}\PH\Pg,\\
                \Pg q       & \to \Pe^+\nu_\Pe \mu^-\bar{\nu}_\mu\Pb\bar{\Pb}\PH q,\\
                \Pg \bar{q} & \to \Pe^+\nu_\Pe \mu^-\bar{\nu}_\mu\Pb\bar{\Pb}\PH \bar{q},
        \end{aligned}
\end{equation}
where the $\Pg\Pg$ process involves 1578 tree diagrams and the $q \bar{q}$, $\Pg 
q$ and $\Pg \bar{q}$ processes, all related by crossing symmetry, 614 tree 
diagrams each. 

Gluon Bremsstrahlung in the real corrections gives rise to IR
divergences by soft or collinear configurations, which cancel for the
final state for infrared-safe observables upon combination with the
virtual corrections. Singularities from collinear initial-state
splitting factorize and can be removed by $\overline{\text{MS}}$
redefinition of the parton distribution functions. We employ the
Catani--Seymour subtraction formalism
\cite{Catani:1996vz,Catani:2002hc} for the regularization and
analytical cancellation of IR singularities.
Both the amplitudes for the real-correction subprocesses as well as the
colour and spin-correlated amplitudes of the subtraction terms
have been calculated with \recola.

\subsection{Virtual corrections}
\label{ssec:VirtualCorrections}

\begin{figure}
        \newcommand{\myframebox}{\framebox}
        \renewcommand{\myframebox}{\relax}
        \setlength{\parskip}{-10pt}
        \captionsetup[subfigure]{margin=0pt}
        \begin{subfigure}{0.32\linewidth}
                \subcaption{}
                \myframebox{
                        \includegraphics[width=\linewidth]{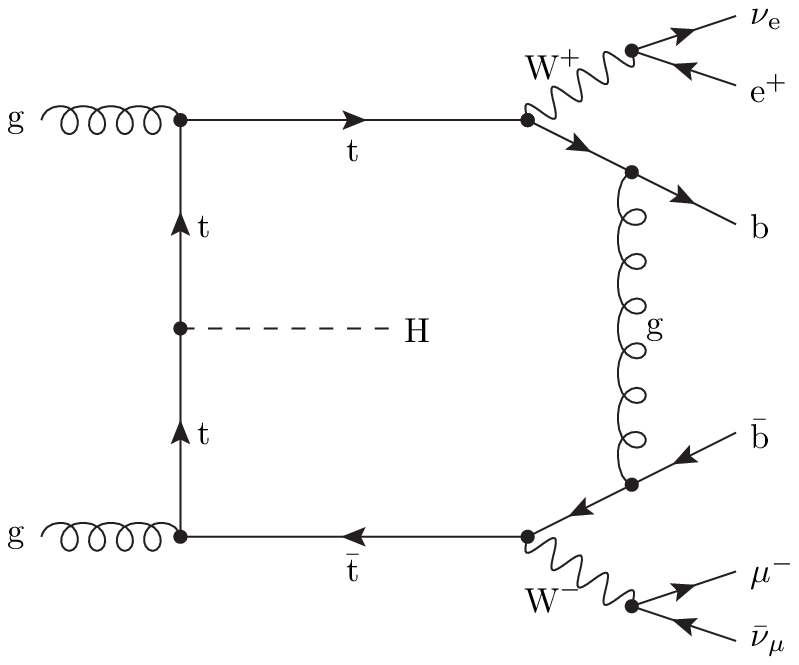}
                }
                \label{fig:virt_2tops_gg_tchannel} 
        \end{subfigure}
        \begin{subfigure}{0.32\linewidth}
                \subcaption{}
                \myframebox{
                        \includegraphics[width=\linewidth]{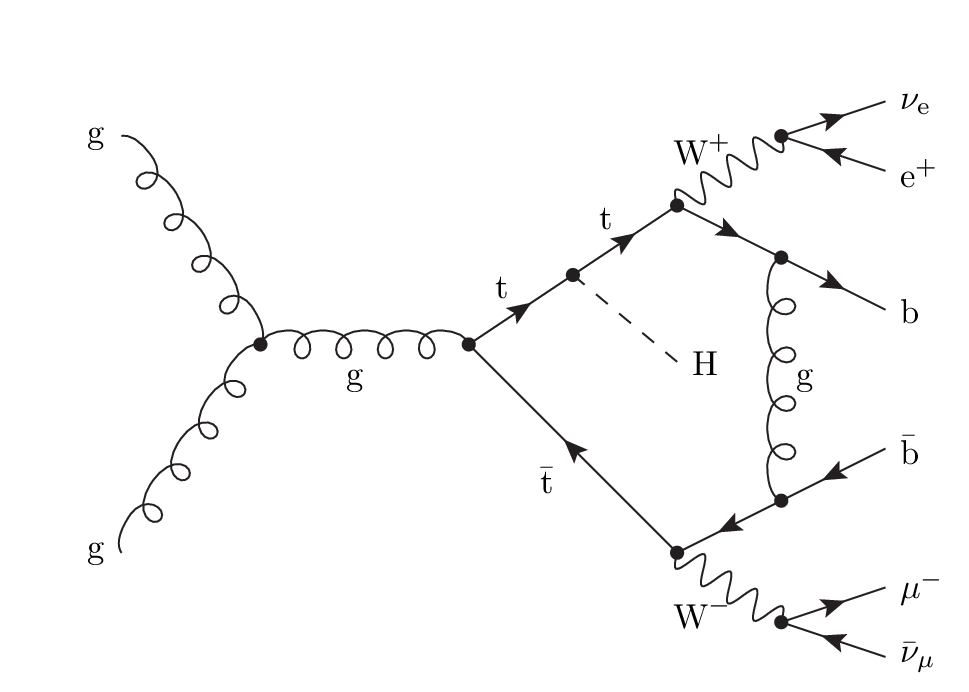}
                }
                \label{fig:virt_2tops_gg_schannel} 
        \end{subfigure}
        \begin{subfigure}{0.32\linewidth}
                \subcaption{}
                \myframebox{
                        \includegraphics[width=\linewidth]{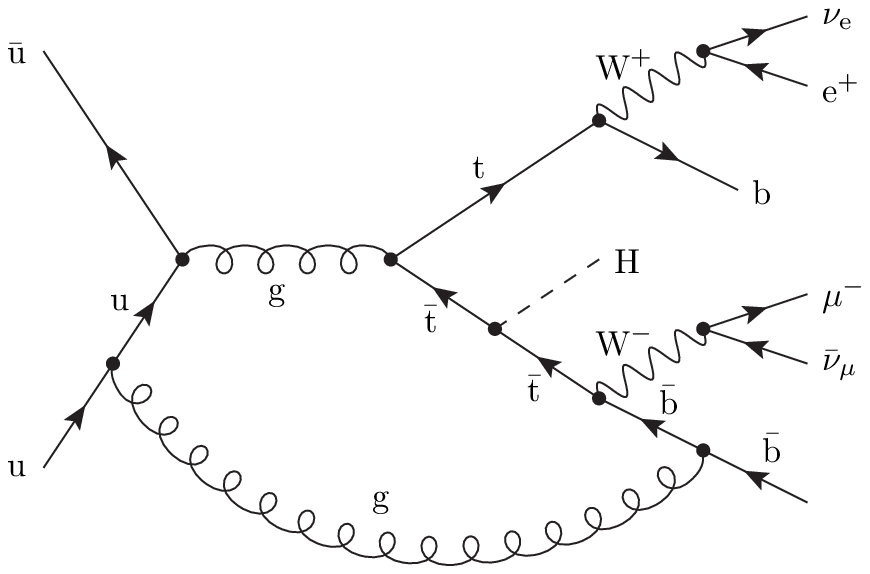}
                }
                \label{fig:virt_2tops_uxu_schannel} 
        \end{subfigure}

        \caption{\label{fig:nlo_loop_feynman_diagrams}%
                Representative hexagon and heptagon one-loop Feynman diagrams with two top-quark resonances.
        }
\end{figure}
The partonic subprocesses for the virtual QCD corrections can be identified with 
those at LO. We compute the virtual corrections in the 't~Hooft--Feynman gauge, 
where the $\Pg\Pg$ process involves 9074 loop diagrams and the $q \bar{q}$ 
processes 2404 loop diagrams each. The most complicated one-loop diagrams are 
heptagons (Sample diagrams are displayed in 
\reffi{fig:nlo_loop_feynman_diagrams}).

The resonant top quarks, Z~bosons and W~bosons are treated in the
complex-mass scheme \cite{Denner:1999gp,Denner:2005fg,Denner:2006ic}, where the masses of unstable
particles are consistently treated as complex quantities leading in
particular to a complex weak mixing angle,
\begin{equation}
\label{e:WZcms}
\mu^2_\PW=\MW^2-\ri\MW\Gamma_\PW,\qquad
\mu^2_\PZ=\MZ^2-\ri\MZ\Gamma_\PZ,\qquad
\cos\theta_{\mathrm{w}}=\frac{\mu_\PW}{\mu_\PZ}.
\end{equation}
For the renormalization we use the on-shell renormalization scheme as
described in \citere{Denner:2005fg} for the complex-mass scheme.

For the computation of the matrix elements for the virtual corrections
we employ \recola~\cite{Actis:2012qn} in dimensional regularisation,
which integrates the \collier~\cite{Denner:2014gla,Collier} library
for the numerical evaluation of one-loop scalar \cite{'tHooft:1978xw,Beenakker:1988jr,Dittmaier:2003bc,Denner:2010tr} and tensor integrals
\cite{Passarino:1978jh,Denner:2002ii,Denner:2005nn}.
We compared our results for the virtual NLO contribution to the
squared amplitude, $2\Re\mathcal{M}^*_0\mathcal{M}_1$, for many
phase-space points with \madgraph~\cite{Alwall:2014hca} (see
\refse{sec:Checks} for details).

\section{Numerical Results}
\label{sec:results}

\subsection{Input parameters, jet definition, cuts and scale choice}
\label{ssec:InputParameters}

We present results for integrated cross sections and differential
distributions for the LHC operating at $\sqrt{s}=13\TeV$. For the
computation of the hadronic cross section we employ LHAPDF 6.05 with
\change{CT10NLO} parton distributions \change{at LO and NLO QCD.} We use the value
of the strong coupling constant $\alphas$ as provided by LHAPDF based
on a one-loop (two-loop) accuracy at LO (NLO) with $N_\text{F}=5$
active flavours. In the renormalization of \alphas the top-quark loop
in the gluon self-energy is subtracted at zero momentum. The running
of \alphas in this scheme is generated by contributions from
light-quark and gluon loops only. For the fixed renormalization
and factorization scale $\mu_\text{fix}=236\GeV$, defined below in
\eqref{eqn:FixedScale}, 
\change{we find
\begin{equation}\label{eqn:Alphas}
    \alphas(\mu_\text{fix}) = 0.103237\ldots\;.
\end{equation}}
We neglect contributions from the suppressed bottom-quark parton
density.

The electromagnetic coupling $\alpha$ is derived from the Fermi constant in the 
$G_\mu$ scheme \cite{Denner:2000bj}, 
\begin{equation}\label{eqn:FermiConstant}
  \alpha = \frac{\sqrt{2}}{\pi} G_\mu \MW^2 \left( 1 - \frac{\MW^2}{\MZ^2} \right),
  \qquad     \GF    = 1.16637\times 10^{-5}\GeV.            
\end{equation}

We compute the width of the top quark \Gt for unstable W bosons and massless 
bottom quarks according to \citere{Jezabek:1988iv}. At NLO QCD it reads
\newcommand{\gammaw}{\ensuremath{\gamma_\text{W}}}
\begin{align}\label{eqn:TopQuarkWidth}
        \Gamma_\Pt^\text{NLO} = \frac{\GF\Mt^5}{16\sqrt{2}\pi^2\MW^2}\int_0^1\!\!
                \frac{\text{d}y\,\gammaw}{\left(1-y/\bar{y}\right)^2+\gammaw^2 } \left(F_0(y)-\frac{2\alphas}{3\pi}F_1(y)\right)
\end{align}
with $\gammaw = \GW/\MW$, $\bar{y} = (\MW/\Mt)^2$, $\alphas=\left.\alphas(\Mt)\right|_\text{NLO}$ and
\begin{align}
        F_0(y) & = 2(1-y)^2(1+2y)\\
        F_1(y) & = 
        \begin{aligned}[t]
                & 2(1-y)^2(1+2y)\left[\pi^2+2\text{Li}_2(y)-2\text{Li}_2(1-y)\right]\\
                           & + 4y(1-y-2y^2)\ln(y)+2(1-y)^2(5+4y)\ln(1-y)\\
                           &-(1-y)(5+9y-6y^2).
        \end{aligned}
\end{align}
For the top-quark width at LO we neglect the correction term $-2\alpha_sF_1(y)/(3\pi)$.

As input, we employ the following numerical values for the masses and widths:
\begin{equation}\label{eqn:ParticleMassesAndWidths}
        \begin{aligned}
                \Mt   &= 173\GeV,           & \Gt^\text{LO}  &= 1.472886\ldots\GeV,\quad & \Gt^\text{NLO} &= 1.346449\ldots\GeV,\\
                \MZOS &=  91.1876\GeV,\quad & \GZOS &= 2.4952\GeV,\\
                \MWOS &=  80.385\GeV,       & \GWOS &= 2.0850\GeV,\\
                \MH   &= 126\GeV,           & \\
        \end{aligned}
\end{equation}
which includes the measured value of the Higgs-boson mass with zero width since 
we assume it to be stable. We neglect the masses and widths of all other 
quarks and leptons.

We convert the measured on-shell values (OS) for the masses and widths of the W 
and Z boson into pole values for the gauge bosons ($V=\PW,\PZ$) according to 
\citere{Bardin:1988xt},
\newcommand{\MVOS}{\ensuremath{M_V^\text{OS}}\xspace}%
\newcommand{\GVOS}{\ensuremath{\Gamma_V^\text{OS}}\xspace}%
\begin{equation}
        M_V = \MVOS/\sqrt{1+(\GVOS/\MVOS)^2}\,,\qquad  \Gamma_V = \GVOS/\sqrt{1+(\GVOS/\MVOS)^2},
\end{equation}
that enter the calculation.

We use the anti-$k_\text{T}$ algorithm \cite{Cacciari:2008gp} for the 
jet reconstruction with a jet-resolution parameter $R=0.4$. The distance 
between two jets $i$ and $j$ in the rapidity--azimuthal plane is defined as
\begin{equation}\label{eqn:DeltaR}
        R_{ij} = \sqrt{(\phi_i-\phi_j)^2+(y_i-y_j)^2},
\end{equation}
with the azimuthal angle $\phi_i$ and the rapidity $y_i=\frac{1}{2}\ln 
\frac{E+p_z}{E-p_z}$ of jet $i$, where $E$ is the energy and $p_z$ the component 
of momentum along the beam axis. Only final-state quarks and gluons with 
rapidity $|y|<5$ are clustered into infrared-safe jets. 

After recombination we impose standard selection cuts on transverse momenta and 
rapidities of charged leptons and b~jets, missing transverse momentum and 
distance between b~jets according to \eqref{eqn:DeltaR}. We require  
two b jets and two charged leptons in the final state, with bottom quarks in jets 
leading to b jets, and
\begin{equation}\label{eqn:cuts}
        \begin{aligned}
                \text{b jets:}                      && \ptsub{\Pb}         &>  25\GeV,  & |y_\Pb|   &< 2.5, &\qquad\\
                \text{charged lepton:}              && \ptsub{\Pl}         &>  20\GeV,  & |y_{\Pl}| &< 2.5, &\\
                \text{missing transverse momentum:} && \ptsub{\text{miss}} &>  20\GeV,                      &\\
                \text{b-jet--b-jet distance:}       && \Delta R_{\Pb\Pb}   &> 0.4.                          &\\
        \end{aligned}
\end{equation}

We have identified the renormalization scale with the factorization scale 
$\mu=\mu_\text{R} = \mu_\text{F}$ and have considered a fixed reference scale 
set to half the partonic threshold energy for $\Pt\bar{\Pt}\PH$ production 
according to \citere{Beenakker:2002nc}:
\begin{equation}\label{eqn:FixedScale}
        \mu_\text{fix} = \mu_\text{R} = \mu_\text{F} = \frac{1}{2}\left(2m_\Pt + m_\PH\right) = 236\GeV.
\end{equation}
Alternatively, we use a dynamical scale following \citere{Frederix:2011zi}
\begin{equation}\label{eqn:DynamicalScale}
        \mu_\text{dyn} = \mu_\text{R} = \mu_\text{F} = \left(m_{\text{T},\Pt}\,m_{\text{T},\bar{\Pt}}\, m_{\text{T},\PH}\right)^\frac{1}{3}\quad\text{with}\quad m_{\text{T}}=\sqrt{m^2+p^2_\text{T}},
\end{equation}
which corresponds to the geometric mean of the top-quark,
antitop-quark and Higgs-boson transverse masses. For the comparison of
the scale choices we compute the logarithmic scale average
$\bar{\mu}_\text{dyn}$ of the dynamical scale, defined as
\begin{equation}\label{eqn:LogarithmicScaleAverage}
        \ln\bar{\mu}_\text{dyn}=\frac{\int\ln(\mu_\text{dyn})\text{d}\sigma}{\int\text{d}\sigma}.
\end{equation}

\change{ The scale uncertainty of the LO and NLO cross section is
  determined by variation of the renormalization and factorization
  scales $\mu_\text{R}$ and $\mu_\text{F}$ around the central value
  $\mu_0=\mu_\text{fix}$ and $\mu_0=\mu_\text{dyn}$ for the fixed and
  dynamical scale choice, respectively.. While varying the
  renormalization scale in PDFs and matrix elements, the top-quark
  width remains fixed as computed at the top-quark mass according to
  \eqref{eqn:TopQuarkWidth}.  For the investigation of the scale
  dependence of the integrated cross section in
  \reffi{plot:scale_dependence} we vary the scale $\mu$ up and down by
  a factor of eight for the LO and NLO integrated cross section for
  the three cases: 1) $\mu_\text{R} = \mu_\text{F}=\mu$, 2)
  $\mu_\text{R} = \mu$, $\mu_\text{F}=\mu_0$, 3) $\mu_\text{F} = \mu$,
  $\mu_\text{R}=\mu_0$. While we show all three cases for the
  dynamical scale, we show only the first case for the fixed scale
  choice in
  \reffi{plot:scale_dependence}.  For all other results, i.e.\ those
  in \refta{table:results_summary} and
  \reffis{fig:differential_distributions_fixed_scale}--\ref{fig:further_differential_distributions},
  the scale uncertainties are determined from factor-two variations as
  follows.  We compute integrated and differential cross sections at
  seven scale pairs, $(\mu_\text{R}/\mu_0$,
  $\mu_\text{F}/\mu_0)=(0.5,0.5),(0.5,1),(1,0.5),(1,1),(1,2),(2,1),(2,2)$.
  The central value corresponds to $(\mu_\text{R}/\mu_0$,
  $\mu_\text{F}/\mu_0)=(1,1)$ and the error band is constructed from
  the envelope of these seven calculations.  }

\subsection{Integrated cross section and scale dependence}
\label{ssec:IntegratedCrossSectionScaleDependence}

\begin{table}
        \centering
        \renewcommand\arraystretch{1.2}
        \rowcolors{2}{tablerowcolor}{}
        \begin{tabular}{lllll}
                \toprule\rowcolor{tableheadcolor}       
                \multicolumn{1}{>{\columncolor{tableheadcolor}}l}{\boldmath$\mu_0$}
                & \multicolumn{1}{>{\columncolor{tableheadcolor}}l}{\textbf{ch.}} 
                & \multicolumn{1}{>{\columncolor{tableheadcolor}}c}{\textbf{\boldmath$\sigma_\textbf{LO}$ [fb]}}%
                & \multicolumn{1}{>{\columncolor{tableheadcolor}}c}{\textbf{\boldmath$\sigma_\textbf{NLO}$ [fb]}}%
                & \multicolumn{1}{>{\columncolor{tableheadcolor}}c}{\boldmath$K$}\\
                \midrule            
                $\mu_\text{dyn}$ & $\Pg \Pg$        &  $1.5906(1) ^{+33.7\%}_{-23.6\%}$ & $2.024(3) ^{+8.4\%  }_{-16.2\%}$   & 1.273(2)\\
                                 & $q \bar{q}$      &  $0.67498(9)^{+24.1\%}_{-18.1\%}$ & $0.495(1) ^{+17.2\%  }_{-39.5\%}$  & 0.733(2)\\
                                 & $\Pg \brabar{q}$ &                                   & $0.136(1) ^{+295\%   }_{-166\%}$ &     \\\cmidrule{2-5}\rowcolor{tableheadcolor}
                                 & $\Pp\Pp$         &  $2.2656(1) ^{+30.8\%}_{-22.0\%}$ & $2.656(3) ^{+0.9\%   }_{-4.6\%}$   & 1.172(1)\\
                \midrule
                $\mu_\text{fix}$ & $\Pg \Pg$        &  $1.5681(1) ^{+33.9\%}_{-23.7\%}$ & $2.011(3) ^{+8.5\%}_{-16.4\%}   $ & 1.282(2) \\
                                 & $q \bar{q}$      &  $0.67199(9)^{+24.2\%}_{-18.2\%}$ & $0.495(1) ^{+17.0\%}_{-39.4\%}  $ & 0.737(2) \\
                                 & $\Pg\brabar{q}$  &                                   & $0.127(1) ^{+310\%}_{-175\%}    $ &     \\\cmidrule{2-5}\rowcolor{tableheadcolor}
                                 & $\Pp\Pp$         &  $2.2401(1) ^{+31.0\%}_{-22.0\%}$ & $2.633(3) ^{+0.6\% }_{-5.0\%}   $ & 1.176(1) \\
                \bottomrule
        \end{tabular}
        \caption[Composition of the integrated cross section]{\label{table:results_summary}
                Composition of the integrated cross section for $\Pp\Pp \to 
                \Pe^+\nu_\Pe \mu^- \bar{\nu}_\mu \Pb \bar{\Pb} \PH(\Pj)$ at the LHC at $\sqrt{s}=13\TeV$ 
                with both the dynamical ($\mu_\text{dyn}$) and the fixed scale 
                ($\mu_\text{fix}$) as denoted in column one. In column two we list the partonic 
                initial states, where $q=\Pu,\Pd,\Pc,\Ps$ and $\brabar{q}=q,\bar{q}$. The third 
                and fourth column give the integrated cross sections in fb for LO and 
                NLO, resp. \change{The upper and lower variations correspond to the envelope of seven scale
                pairs $(\mu_\text{R}/\mu_0$, $\mu_\text{F}/\mu_0)=(0.5,0.5),(0.5,1),(1,0.5),(1,1),(1,2),(2,1),(2,2)$.}
                The last column provides the $K$ 
                factor with $K=\sigma_\text{NLO}/\sigma_\text{LO}$.
        }
\end{table}

In \refta{table:results_summary} we present the integrated cross
sections with fixed \eqref{eqn:FixedScale} and dynamical scale
\eqref{eqn:DynamicalScale} at the LHC at $\sqrt{s}=13\TeV$
corresponding to the input parameters \eqref{eqn:Alphas},
\eqref{eqn:FermiConstant} and \eqref{eqn:ParticleMassesAndWidths} and
the cuts as defined in \eqref{eqn:cuts}.  The results include only
contributions of $\order{\alphas\alpha^{5/2}}$ for LO amplitudes and
the corresponding ${\cal O}(\alphas)$ QCD corrections. We neglect
possible contributions to $q \bar{q}$ processes of
$\order{\alpha^{7/2}}$ for LO amplitudes, which we determined to be
about 2 per mille of the integrated cross section at LO for the setup
described above. \change{We also do not include partonic channels with
  incoming bottom quarks. At LO and using the dynamical
  scale \eqref{eqn:DynamicalScale}, these contribute
  $4.869(1)\times10^{-3}\,\mathrm{fb}$, i.e.\ $0.2\%$, to the
  integrated cross section. Moreover, we calculated the integrated
  cross section at LO with finite bottom-quark masses, leading to a
  reduction of the cross section for our set of cuts by 0.03\%.}

The use of the dynamical scale instead of the fixed scale increases
the LO and NLO cross sections by only about 1\,\%, and the $K$~factor
is \change{1.172 for the dynamical scale and 1.176 for the fixed scale.}
The similar quality of both scale choices is also
supported by the logarithmic scale average of the dynamical scale as
defined in \eqref{eqn:LogarithmicScaleAverage}. With
\change{$\bar{\mu}_\text{dyn}=222.3\GeV$} it corresponds to a slight effective
decrease of the fixed scale by only about 6\,\%.

Since integrated cross sections and NLO effects are very similar, the
following considerations hold true for both, the dynamical and fixed
scale: The major contributions to the cross section originate from the
gluon-fusion process, with about 70\,\% at LO while increasing at NLO
to 76\,\%. The contribution of the quark--antiquark annihilation drops
from about 30\,\% at LO to 19\,\%. At NLO the gluon--(anti)quark
induced real-radiation subprocesses contribute about 5\,\% to the
integrated cross section. The inclusion of NLO QCD corrections reduces
the scale dependence from 31\,\% to 5\,\%.

\begin{figure}
        \begin{minipage}[t]{.490\textwidth}
                \includegraphics[width=\linewidth]{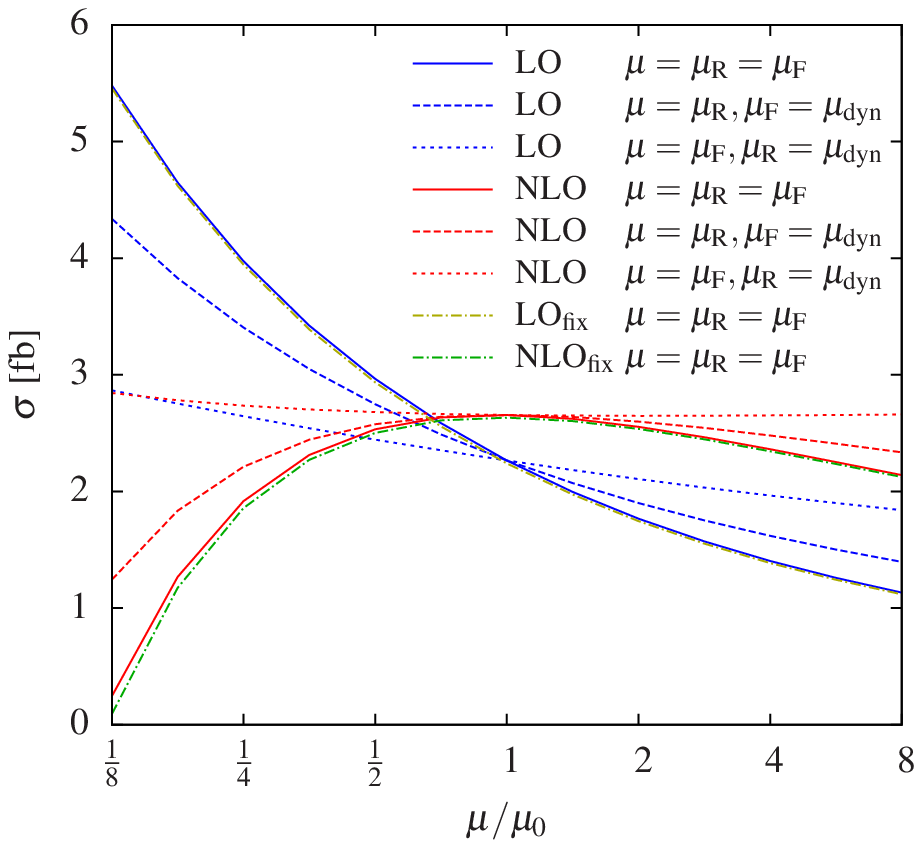}
                \caption{\label{plot:scale_dependence}%
                  Scale dependence of the LO and NLO integrated cross
                  section at the $13\TeV$ LHC. \change{The
                    renormalization and factorization scales are
                    varied around the central values of the fixed
                    ($\mu_0=\mu_\text{fix}$, dash-dotted lines) and
                    dynamical scale ($\mu_0=\mu_\text{dyn}$, solid
                    lines). For the dynamical scale the variation with
                    $\mu_\text{R}$ while keeping
                    $\mu_\text{F}=\mu_\text{dyn}$ fixed and vice versa
                    is shown with dashed lines.}}
        \end{minipage}
        \hfill
        \begin{minipage}[t]{.490\textwidth}
                \includegraphics[width=\linewidth]{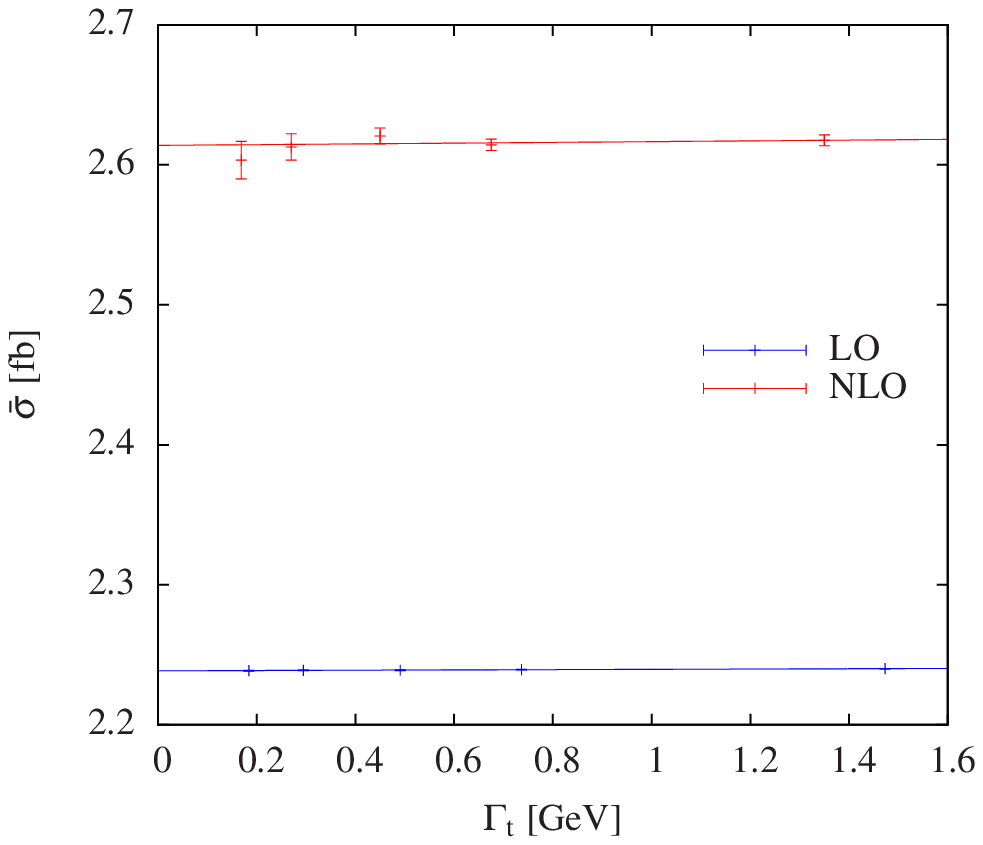}
                \caption{\label{plot:top_width_extrapolation}%
                  Zero-top-width extrapolation of the LO and NLO cross
                  section at the LHC at $\sqrt{s}=13\TeV$ for fixed
                  scale $\mu_0=\mu_\text{fix}$.}
        \end{minipage}
\end{figure}

\change{
  We display the dependence of the integrated LO (blue) and NLO (red)
  cross sections on the values of the fixed and dynamical scale in
  \reffi{plot:scale_dependence}. Solid lines for the dynamical scale
  and dash-dotted lines for the fixed scale show the scale dependence
  for a simultaneous variation of the renormalization and
  factorization scales and dashed lines the individual variation, where
  one of the scales is kept fix at the central value, for the
  dynamical scale only. While the largest scale variation is obtained
  when both scales are changed simultaneously, the smallest effect
  results if only the factorization scale is varied. The cross
  sections for the fixed and dynamical scale choices} are uniformly
shifted relative to each other by about 1\,\% as for the central scale
$\mu_0$ both for LO and NLO except for $\mu<\mu_0/2$, where the fixed
scale leads to a faster decrease of the cross section with $\mu$ as
the dynamical scale. For the fixed and dynamical scale the maximum of
the NLO cross section is near $\mu\simeq\mu_0$, justifying the use of
both scale choices to be stable against scale variations. The
$K$~factor equals one at the slightly lower scale of about $\mu\simeq
0.7\mu_0$

\subsection{Limit of on-shell top quarks}

To determine the effects of non-resonant and off-shell top-quark
contributions on the integrated cross section we perform a numerical
extrapolation to the zero-top-width limit, $\Gamma_\Pt \to 0$. To
this end we plot
\begin{equation}\label{eqn:cross_section_top_quark_extrapolation}
        \bar{\sigma}^{\text{LO/NLO}}(\Gamma_\Pt)=\sigma^{\text{LO/NLO}}(\Gamma_\Pt)\left(\frac{\Gamma_\Pt}{\Gamma_\Pt^{\text{LO/NLO}}}\right)^2
\end{equation}
in the range $0\le \Gamma_\Pt \leq \Gamma_\Pt^{\text{LO/NLO}}$, where
$\Gamma_\Pt^{\text{LO/NLO}}$ is the top-quark width at LO and NLO,
resp., and extrapolate linearly to $\Gamma_\Pt \to 0$, using a linear
regression based on the computed LO and NLO integrated cross sections,
as shown in \reffi{plot:top_width_extrapolation}. The factor
$(\Gamma_\Pt/\Gamma_\Pt^{\text{LO/NLO}})^2$ restores the physical
top-decay branching fraction. Finite-top-width effects can be
extracted by comparing the results for
$\bar{\sigma}^{\text{LO/NLO}}(\Gamma_\Pt\to 0)$ to
$\sigma^{\text{LO/NLO}}(\Gamma_\Pt^{\text{LO/NLO}})$. At the LHC at
$\sqrt{s}=13\TeV$ for fixed scale $\mu_0=\mu_\text{fix}$
finite-top-width effects shift the LO and NLO cross section by
\change{$-0.07\pm0.01\,\%$} and $-0.14\pm0.22\,\%$, respectively, which are
within the expected order of $\Gamma_\Pt/m_\Pt$.  \change{The strong
  suppression of finite-top-width effects is related to the
  requirement of a final state with two hard $\Pb$~jets.  Finite-width
  effects are much more sizable in calculations where phase-space
  regions allowing for associated single-top plus W-boson production
  are included \cite{Cascioli:2013wga}. Such calculations require
  massive bottom quarks to regularize collinear singularities.}

\subsection{Differential distributions}
\label{ssec:DifferentialDistributions}

\begin{figure}
        \setlength{\parskip}{-10pt}
        \begin{subfigure}{0.50\textwidth}
                \subcaption{}
                \includegraphics[width=\textwidth]{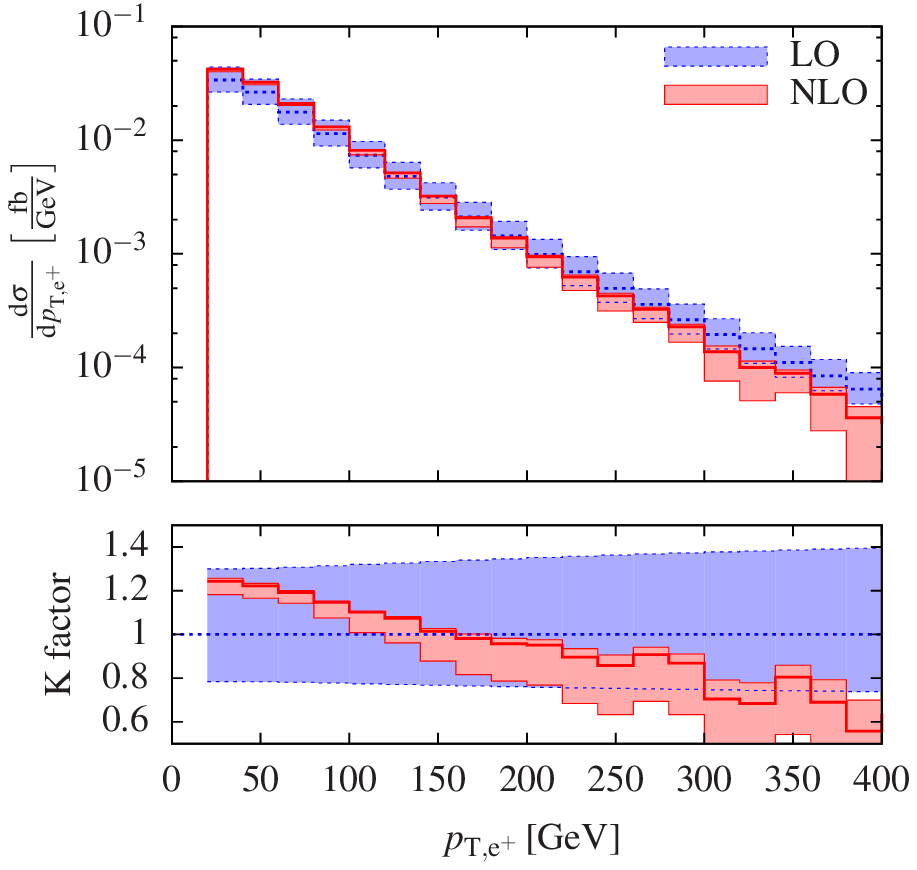}
                \label{plot:transverse_momentum_positron_fix}
        \end{subfigure}
         \hfill
         \begin{subfigure}{0.50\textwidth}
                \subcaption{}
                \includegraphics[width=\textwidth]{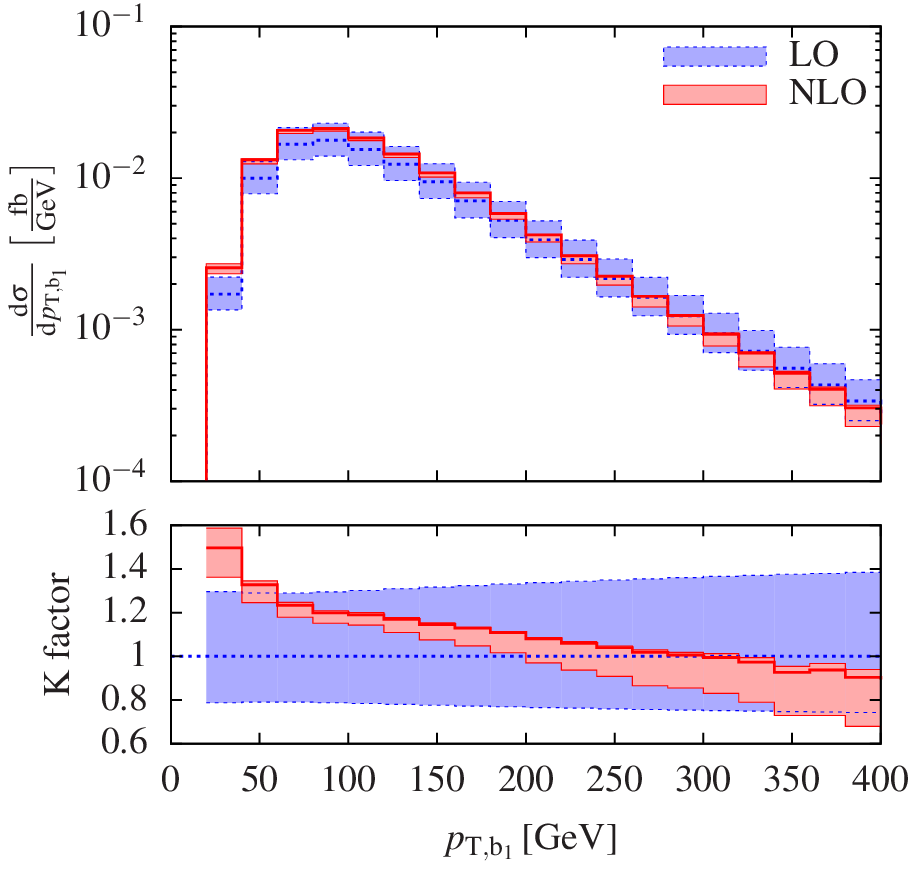}
                \label{plot:transverse_momentum_b1_fix}
        \end{subfigure}
        \vspace*{-3ex}
        \caption{\label{fig:differential_distributions_fixed_scale}%
                Transverse-momentum distributions at the LHC at $\sqrt{s}=13\TeV$ for fixed scale $\mu_0=\mu_\text{fix}$: 
                \subref{plot:transverse_momentum_positron_fix} for the positron~(left) and  %
                \subref{plot:transverse_momentum_b1_fix} for the harder \Pb~jet~(right). %
                The lower panels show the $K$~factor.}
\end{figure}%

\begin{figure}
        \setlength{\parskip}{-10pt}
        \begin{subfigure}{0.50\textwidth}
                \subcaption{}
                \includegraphics[width=\textwidth]{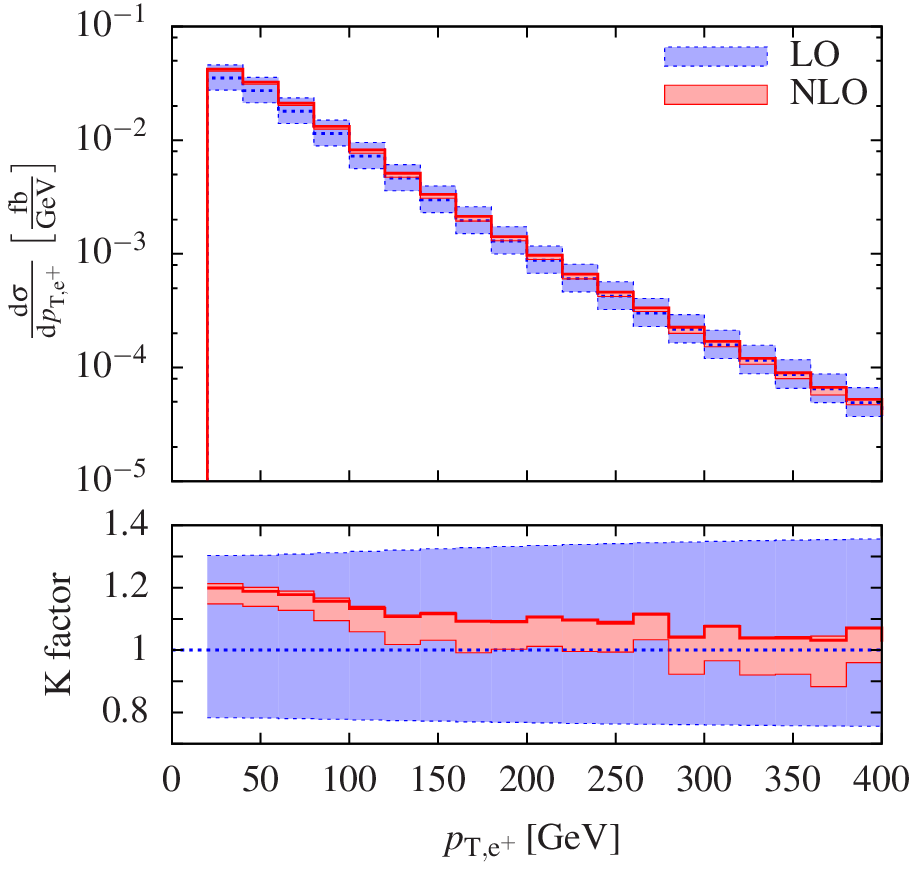}
                \label{plot:transverse_momentum_positron_dyn}
        \end{subfigure}
        \hfill
        \begin{subfigure}{0.50\textwidth}
                \subcaption{}
                \includegraphics[width=\textwidth]{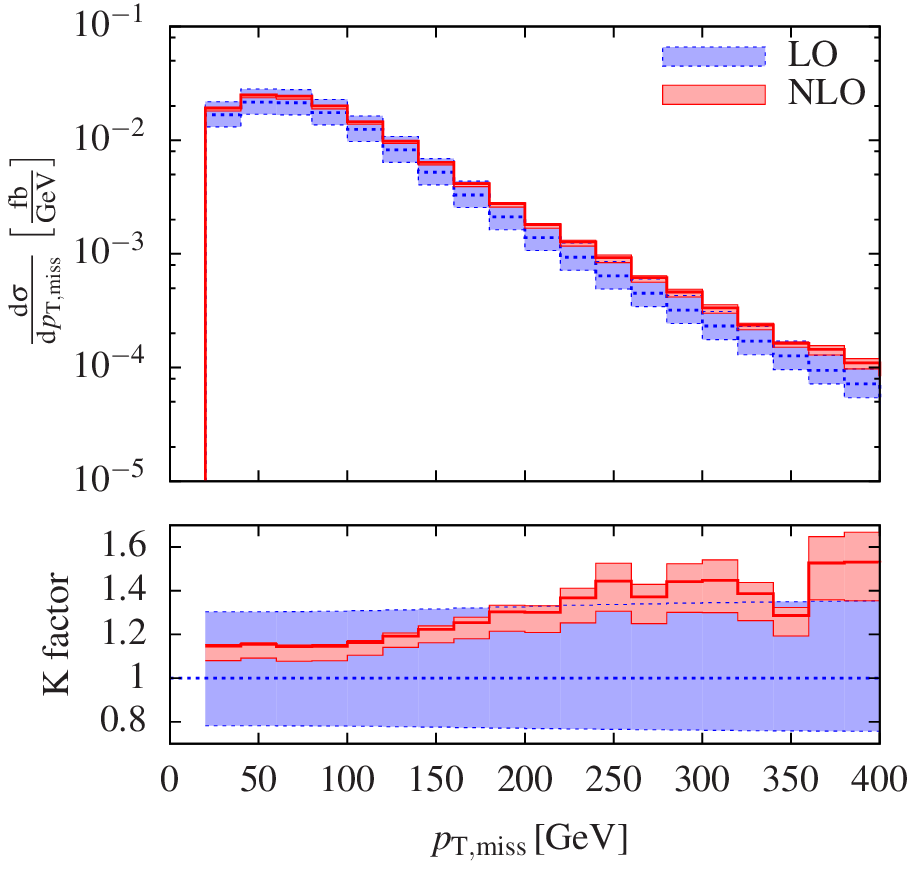}
                \label{plot:transverse_momentum_truth_missing_dyn} 
        \end{subfigure}
        
        \begin{subfigure}{0.50\textwidth}
                \subcaption{}
                \includegraphics[width=\textwidth]{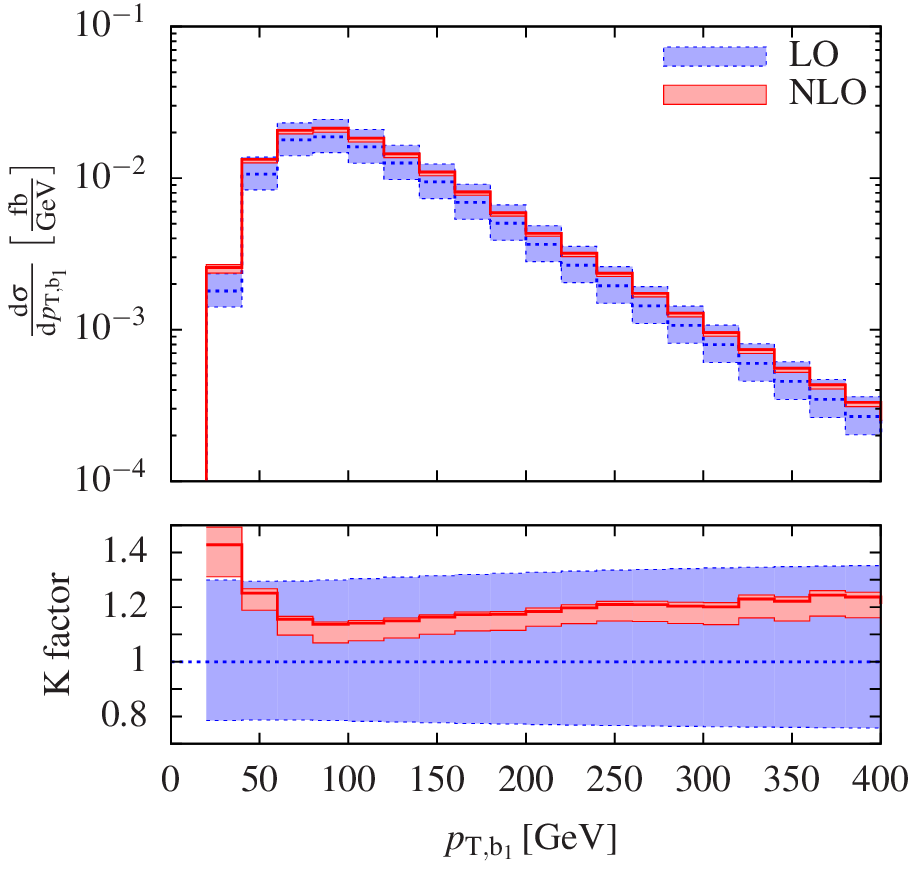}
                \label{plot:transverse_momentum_b1_dyn}
        \end{subfigure}
        \hfill
        \begin{subfigure}{0.50\textwidth}
                \subcaption{}
                \includegraphics[width=\textwidth]{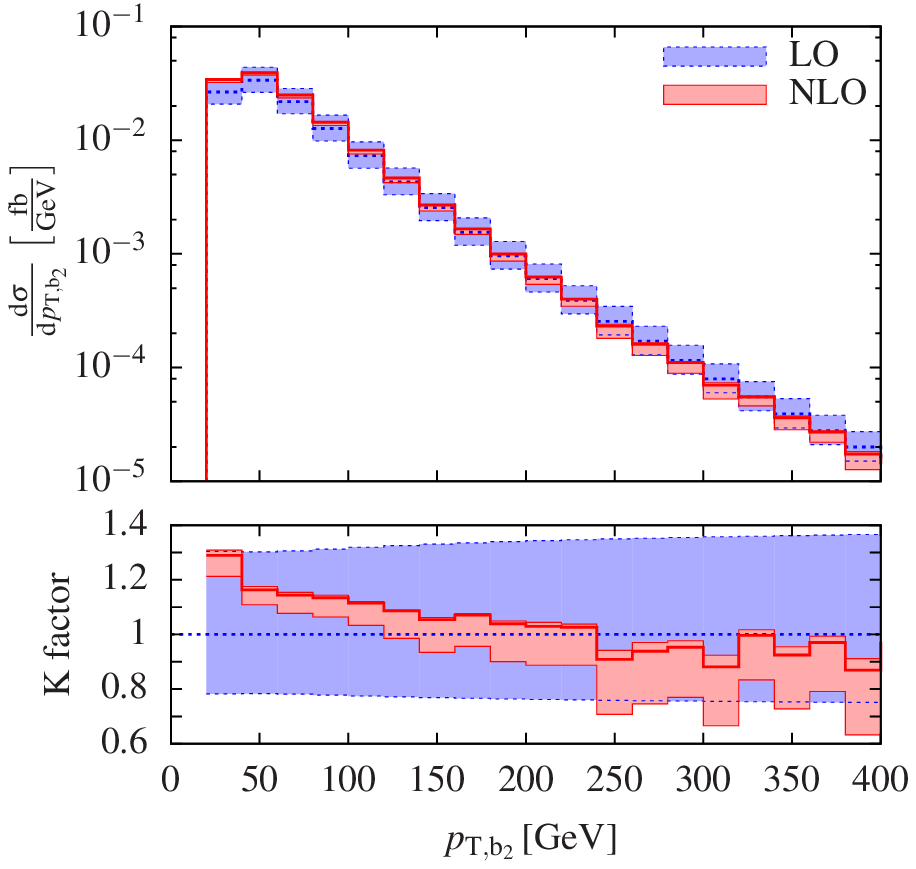}
                \label{plot:transverse_momentum_b2_dyn}
        \end{subfigure}
        
        \begin{subfigure}{0.5\textwidth}
                \subcaption{}
                \includegraphics[width=\textwidth]{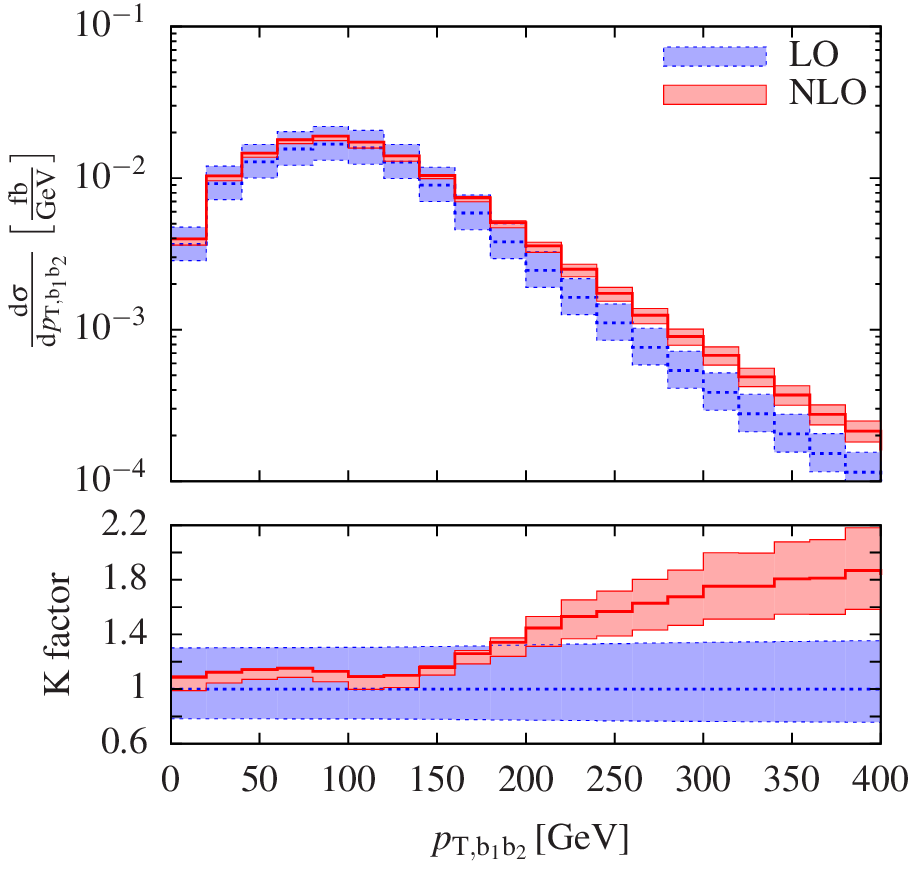}
                \label{plot:transverse_momentum_bb12_dyn}
        \end{subfigure}
        \hfill
        \begin{subfigure}{0.5\textwidth}
                \subcaption{}
                \includegraphics[width=\textwidth]{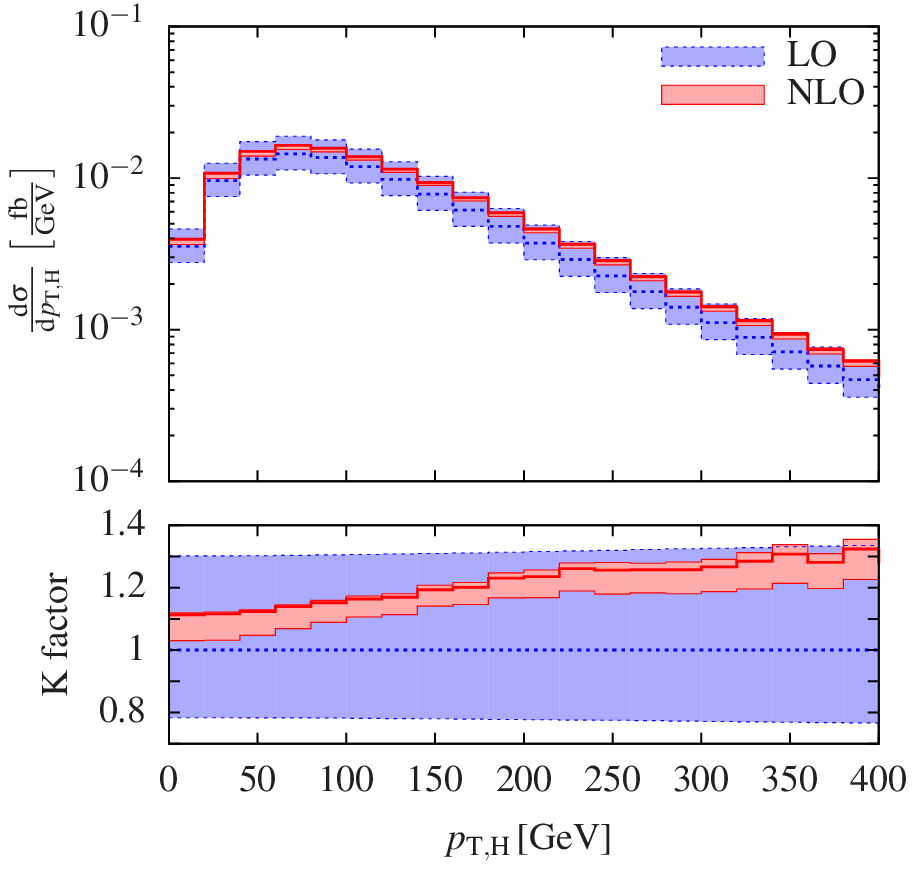}
                \label{plot:transverse_momentum_higgs_dyn}
        \end{subfigure}%
        \vspace*{-3ex}
        \caption{\label{fig:transverse_momentum_distributions}%
                Transverse-momentum distributions at the LHC at $\sqrt{s}=13\TeV$ for dynamical scale $\mu_0=\mu_\text{dyn}$: 
                \subref{plot:transverse_momentum_positron_dyn} for the positron~(upper left), %
                \subref{plot:transverse_momentum_truth_missing_dyn} for missing energy~(upper right), %
                \subref{plot:transverse_momentum_b1_dyn} for the harder \Pb~jet~(middle left), %
                \subref{plot:transverse_momentum_b2_dyn} for the softer \Pb~jet~(middle right), %
                \subref{plot:transverse_momentum_bb12_dyn} for the \Pb-jet pair~(lower left) and %
                \subref{plot:transverse_momentum_higgs_dyn} for the Higgs boson~(lower right). %
                The lower panels show the $K$~factor.}
\end{figure}%

\begin{figure}
        \setlength{\parskip}{-10pt}
        \begin{subfigure}{0.5\textwidth}
                \subcaption{}
                \includegraphics[width=\textwidth]{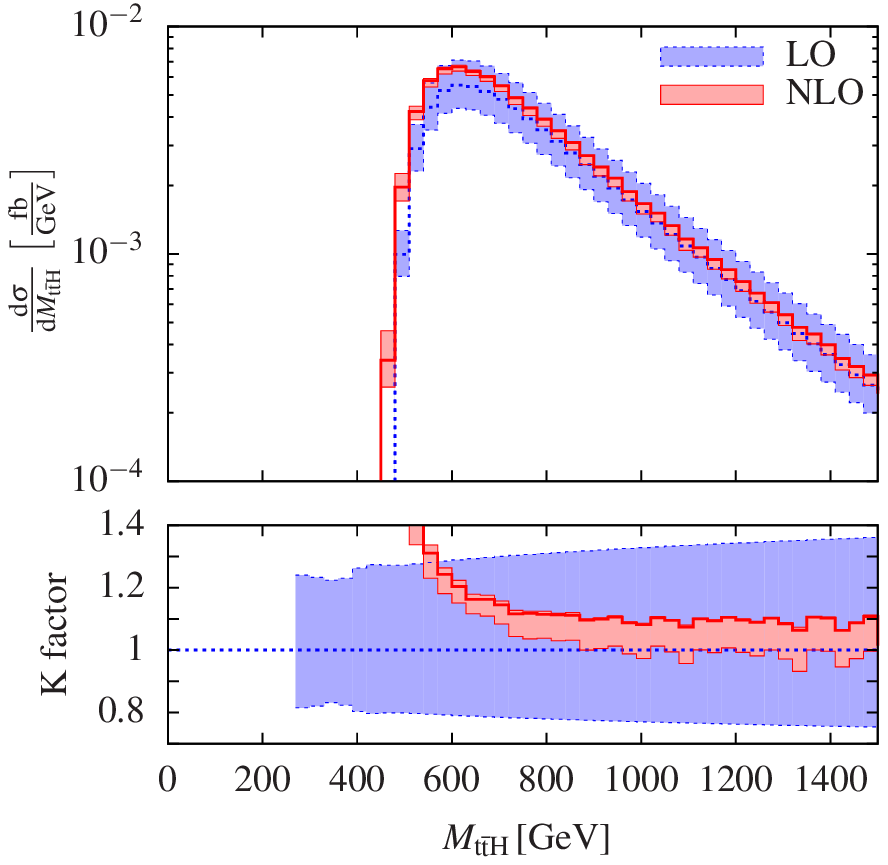}
                \label{plot:invariant_mass_truth_ttxh}
        \end{subfigure}
        \hfill
        \begin{subfigure}{0.5\textwidth}
                \subcaption{}
                \includegraphics[width=\textwidth]{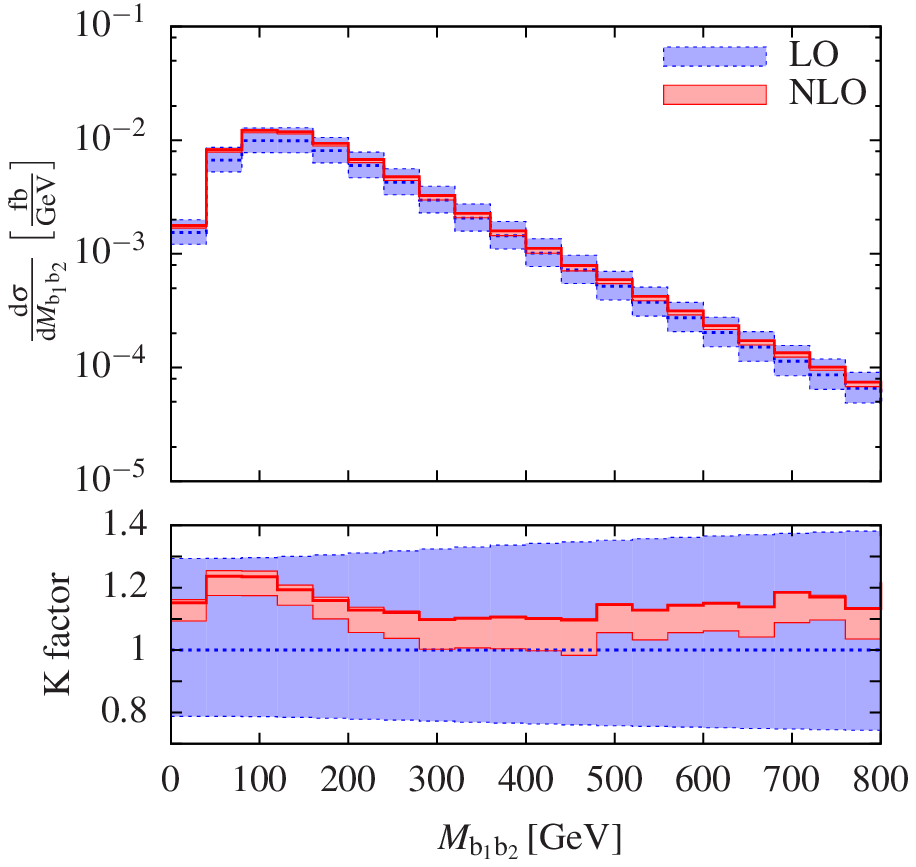}
                \label{plot:invariant_mass_bb12} 
        \end{subfigure}
        
        \begin{subfigure}{0.5\textwidth}
                \subcaption{}
                \includegraphics[width=\textwidth]{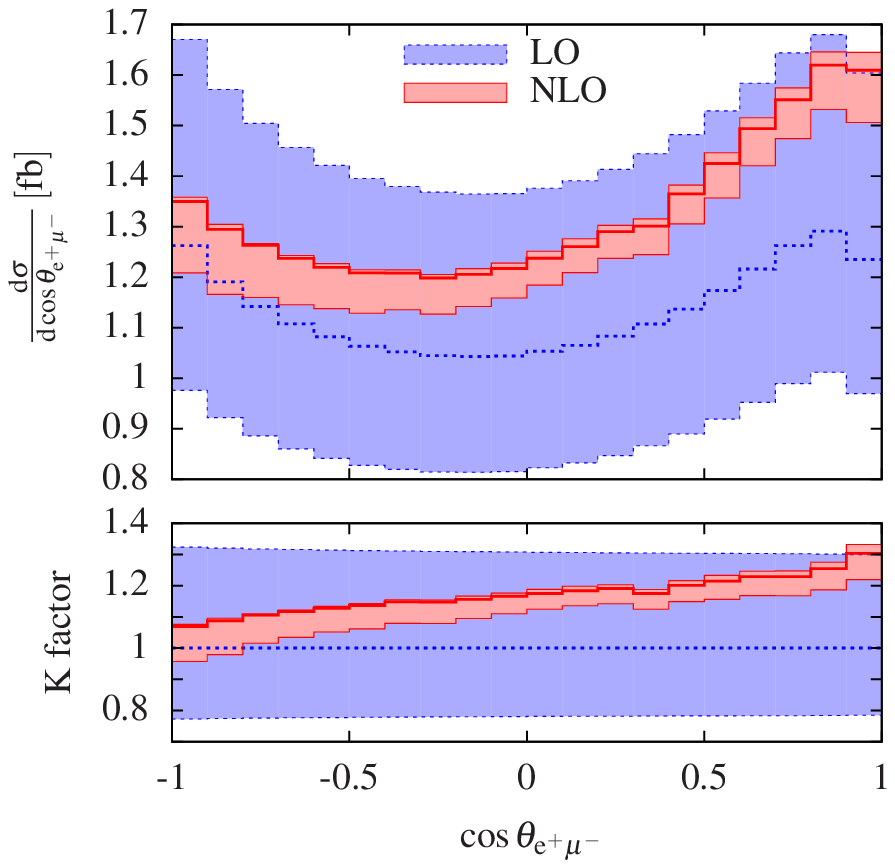}
                \label{plot:cosine_angle_separation_epmu}
        \end{subfigure}
        \hfill
        \begin{subfigure}{0.5\textwidth}
                \subcaption{}
                \includegraphics[width=\textwidth]{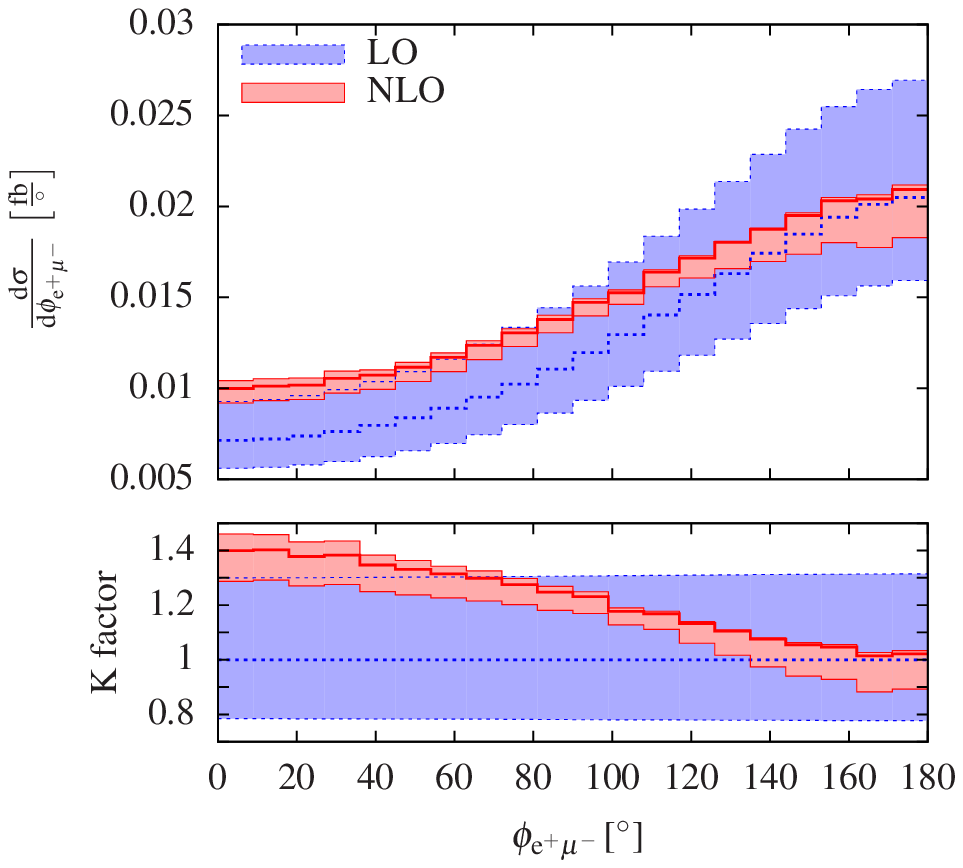}
                \label{plot:azimuthal_angle_separation_epmu}
        \end{subfigure}
        
        \begin{subfigure}{0.5\textwidth}
                \subcaption{}
                \includegraphics[width=\textwidth]{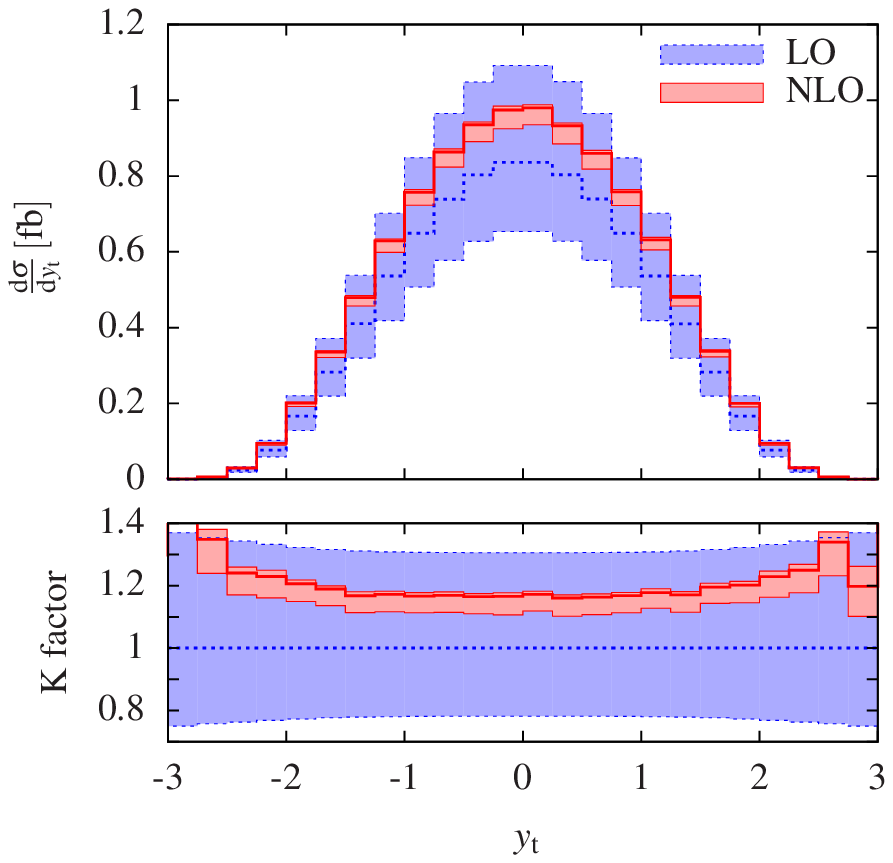}
                \label{plot:rapidity_truth_top}
        \end{subfigure}
        \hfill
        \begin{subfigure}{0.5\textwidth} 
                \subcaption{}
                \includegraphics[width=\textwidth]{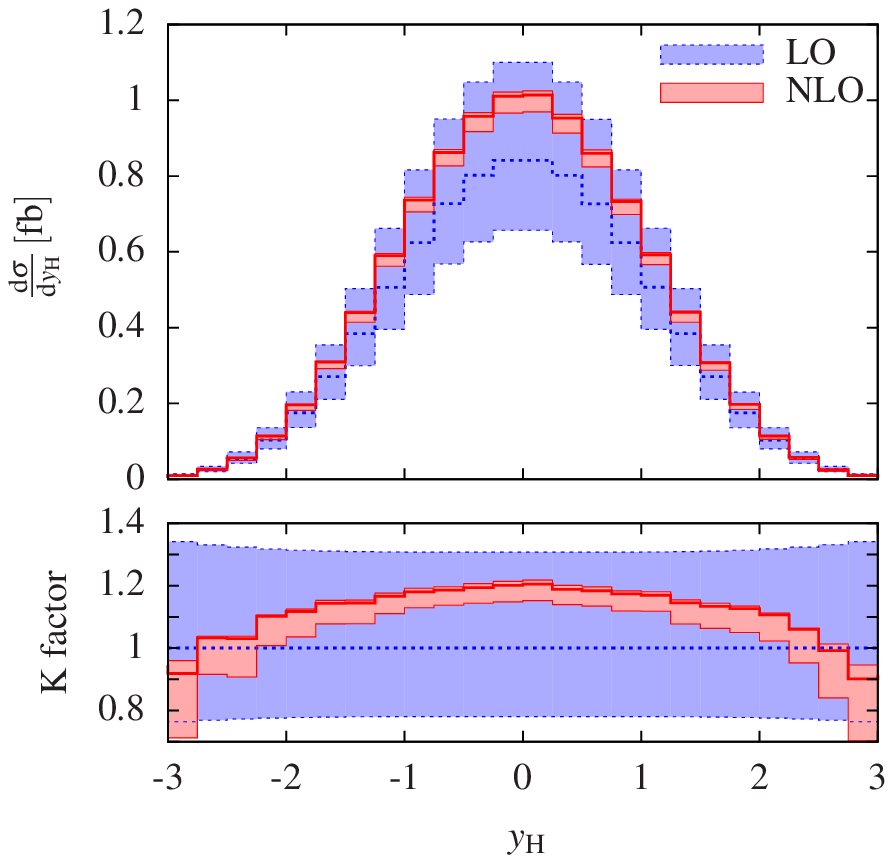}
                \label{plot:rapidity_higgs}
        \end{subfigure}%
        \vspace*{-3ex}
        \caption{\label{fig:further_differential_distributions}%
                Differential distributions at the LHC at $\sqrt{s}=13\TeV$ for dynamical scale $\mu_0=\mu_\text{dyn}$: 
                invariant mass of \subref{plot:invariant_mass_truth_ttxh} the $\Pt\bar{\Pt}\PH$ system~(upper left) and %
                \subref{plot:invariant_mass_bb12} the \Pb-jet pair~(upper right), %
                \subref{plot:cosine_angle_separation_epmu} the cosine of the angle between the positron and the muon~(middle left), %
                \subref{plot:azimuthal_angle_separation_epmu} the azimuthal angle between the positron and the muon in the transverse plane~(middle right), %
                the rapidity of \subref{plot:rapidity_truth_top} the top quark~(lower left) and  %
                of \subref{plot:rapidity_higgs} the Higgs boson~(lower right). %
                The lower panels show the $K$~factor.}
\end{figure}%

In this section we present various differential distributions with two
plots for each observable: The upper plot showing the LO (blue,
dashed) and NLO (red, solid) predictions with uncertainty bands \change{from
the envelope of scale variations by seven pairs,
$(\mu_\text{R}/\mu_0$,$\mu_\text{F}/\mu_0)=(0.5,0.5),(0.5,1),(1,0.5),
(1,1),(1,2),(2,1),(2,2)$.} The lower plot
displays the LO (blue) and NLO (red) predictions normalized to the LO
results at the central scale, i.e.
$K_\text{LO}=\text{d}\sigma_\text{LO}(\mu)/
\text{d}\sigma_\text{LO}(\mu_0)$ and
$K_\text{NLO}=\text{d}\sigma_\text{NLO}(\mu)/
\text{d}\sigma_\text{LO}(\mu_0)$.  Thus, the central red curve
corresponds to the usual NLO correction factor ($K$ factor), defined
as $K=\sigma_\text{NLO}(\mu_0)/\sigma_\text{LO}(\mu_0)$. The blue band
shows the relative scale uncertainty of the LO differential cross
section.

Most of the displayed differential distributions were obtained using
the dynamical scale $\mu_\text{dyn}$, except for
\reffi{fig:differential_distributions_fixed_scale} which illustrates
the effect of the fixed-scale choice on transverse-momentum
distributions.

Where appropriate we compare NLO effects in differential distributions
to those of the related process $\Pp\Pp\to\Pe^+\nu_\Pe
\mu^-\bar{\nu}_\mu\Pb\bar{\Pb}$ presented in \citere{Denner:2012yc}.
In general we find similar NLO effects for most of the distributions,
but being often more distinct for $\Pp\Pp\to\Pe^+\nu_\Pe
\mu^-\bar{\nu}_\mu\Pb\bar{\Pb}$ in \citere{Denner:2012yc}.

Figures~\ref{plot:transverse_momentum_positron_fix} and
\ref{plot:transverse_momentum_b1_fix} display the transverse-momentum
distributions of the positron and the harder \Pb~jet, resp., for the
fixed scale $\mu_0=\mu_\text{fix}$. The $K$~factor of the
transverse-momentum of the positron drops by about 50\,\% within the
plotted range. In the high-\pt tail the NLO predictions move outside
the LO band with a scale variation of almost the same size as the LO
one. The $K$~factor of the transverse-momentum of the harder \Pb~jet
exhibits the same tendency, but not as drastic as for the positron.
It moves outside the LO band for $\pt<60\GeV$ with larger
scale variation as at the average \pt of around 90\GeV.

In \reffi{fig:transverse_momentum_distributions} we collect several
transverse momentum distributions obtained using the dynamical scale
$\mu_0=\mu_\text{dyn}$.
Figures~\ref{plot:transverse_momentum_positron_dyn} and
\ref{plot:transverse_momentum_b1_dyn} show the transverse-momentum
distributions of the positron and the harder \Pb~jet, resp., to
compare with the fixed-scale distributions in
Figures~\ref{plot:transverse_momentum_positron_fix} and
\ref{plot:transverse_momentum_b1_fix} described above: They show
clearly that the dynamical-scale choice improves the perturbative
stability. The $K$~factor changes only slightly (within 20\,\%) over
the displayed range, and the NLO band lies within the LO band.
The residual scale variation is at the level of $10\,\%$ at NLO.

In \reffi{plot:transverse_momentum_truth_missing_dyn} the distribution of 
missing transverse momentum, defined as 
$\ptsub{\text{miss}}=\abs{\vec{p}_{\text{T},\nu_{\Pe}}+\vec{p}_{\text{T},\bar{
\nu}_{\mu}}}$, is shown. The $K$~factor rises for $\ptsub{\text{miss}}\gtrsim
100\GeV$ up to about \change{1.5}.

Figures~\ref{plot:transverse_momentum_b1_dyn} and
\ref{plot:transverse_momentum_b2_dyn} display the distribution of the
transverse momentum of the harder and softer \Pb~jet, resp. While the
$K$~factor for the harder \Pb~jet exhibits a minimum at the maximum of
the distribution and slightly rises towards its tail, the $K$~factor
of the softer \Pb~jet decreases by about 30\,\% in the plotted range.

The distribution of the transverse momentum of the \Pb-jet pair in
Fig.\ 14 of \citere{Denner:2012yc} exhibits a strong suppression of
the $\Pt\bar\Pt$ cross section at LO above
$\ptsub{\Pb\bar{\Pb}}\gtrsim 150\GeV$. This is due to the fact that in
narrow-top-width approximation the \Pb quarks are boosted via their
parent top and antitop quark, which have opposite transverse momenta
resulting in a suppression of a $\Pb\bar\Pb$ system with high \pt at
LO.  The lesser stringent kinematical constraints at NLO result in an
enhancement of the cross section and thus a large $K$~factor for high
\ptsub{\Pb\bar\Pb}. For the $\Pt\bar{\Pt}\PH$ production at hand a
Higgs boson acquiring transverse momentum softens the kinematical
constraint already at LO leading to smaller NLO corrections for high
\ptsub{\Pb\bar\Pb}, which can be seen in
\reffi{plot:transverse_momentum_bb12_dyn}. Furthermore, the $K$~factor
of the distribution resembles the $K$~factor of the missing transverse
momentum due to a kinematically similar configuration, but with a
stronger increase to a value of 1.8 at $\pt\simeq400\GeV$.

\reffi{plot:transverse_momentum_higgs_dyn} displays the
transverse-momentum distribution of the Higgs boson. The average \pt
of the Higgs boson is around 70\GeV. The cross section decreases more
moderately with $\pt$ in the plotted range than for other transverse
momentum distributions presented in
\reffi{fig:transverse_momentum_distributions}.

In \reffi{fig:further_differential_distributions} we present further
differential distributions for other types of observables: the
invariant mass of the $\Pt\bar{\Pt}\PH$ and $\Pb_1\Pb_2$ system in
\reffi{plot:invariant_mass_truth_ttxh} and
\reffi{plot:invariant_mass_bb12}, resp., the cosine of the angle
between the two charged leptons in
\reffi{plot:cosine_angle_separation_epmu}, the azimuthal angle in the
transverse plane between them in
\reffi{plot:azimuthal_angle_separation_epmu}, and the rapidity of the
top quark and the Higgs boson in \reffi{plot:rapidity_truth_top} and
\reffi{plot:rapidity_higgs}, resp. Large NLO corrections appear in the
$M_{\Pt\bar{\Pt}\PH}$ distribution below the $\Pt\bar{\Pt}\PH$
threshold. These arise dominantly from real gluon bremsstrahlung
contributions, where the emitted gluon is not recombined with the
bottom quarks and thus does not contribute to $M_{\Pt\bar{\Pt}\PH}$.
The distribution of the azimuthal angle in the transverse
plane between the two charged leptons exhibits sizeable NLO effects
for small angles similarly as the distribution in the cosine of the angle
between the two charged leptons. The $K$ factor varies by $40\,\%$ for
these distributions.
The rapidity distribution of the Higgs boson features about the
typical NLO effects in the central detector region, which disappear
going into forward or backward direction. The $K$~factor of the
top-quark rapidity distribution on the other hand is almost flat at
the level of the NLO effects of the integrated cross section.

\section{Checks}
\label{sec:Checks}

\begin{figure}
        \begin{subfigure}{0.47\textwidth}
                \captionsetup{skip=-10pt,margin=0pt}
                \subcaption{}
                \ \includegraphics[width=\textwidth]{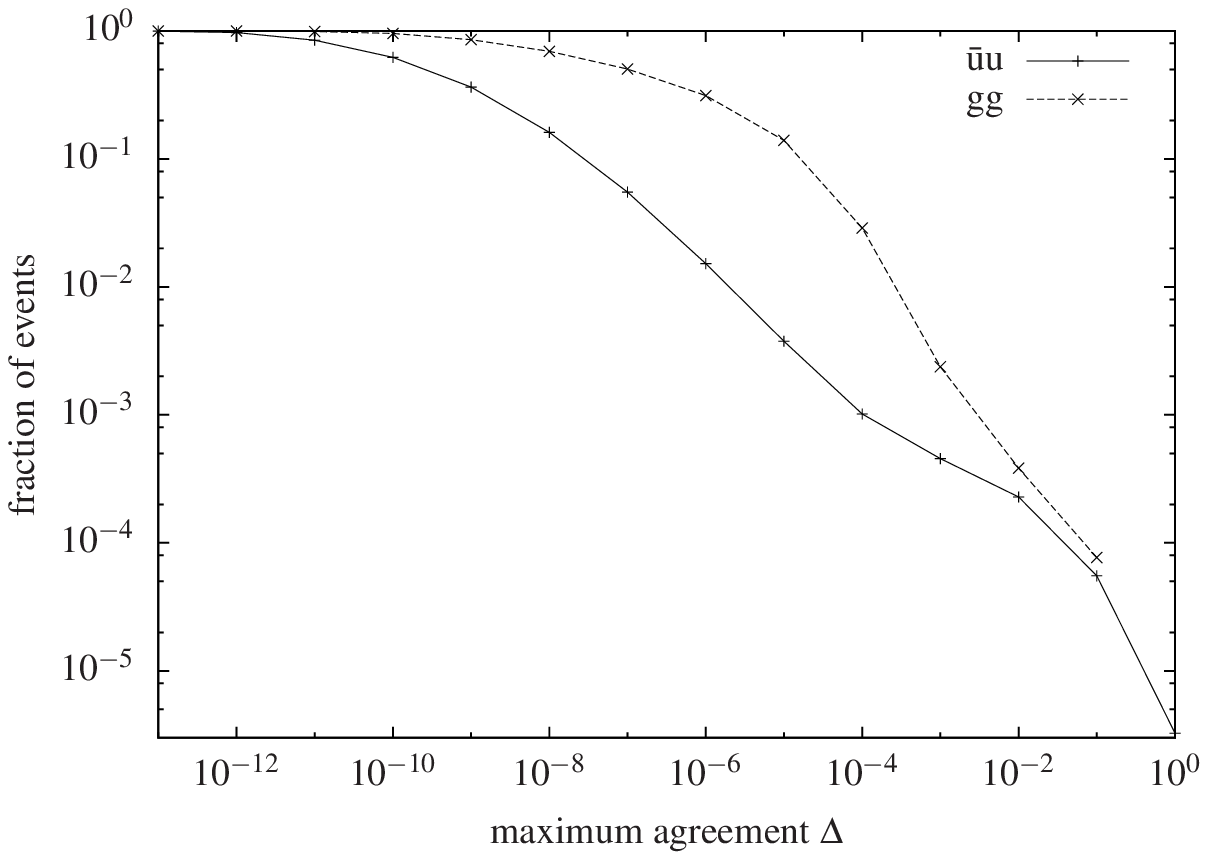}
                \label{plot:mg5_rcl_nlo_agreement}
        \end{subfigure}
        \hfill
        \begin{subfigure}{0.47\textwidth}
                \captionsetup{skip=-10pt,margin=0pt}
                \subcaption{}
                \ \includegraphics[width=\textwidth]{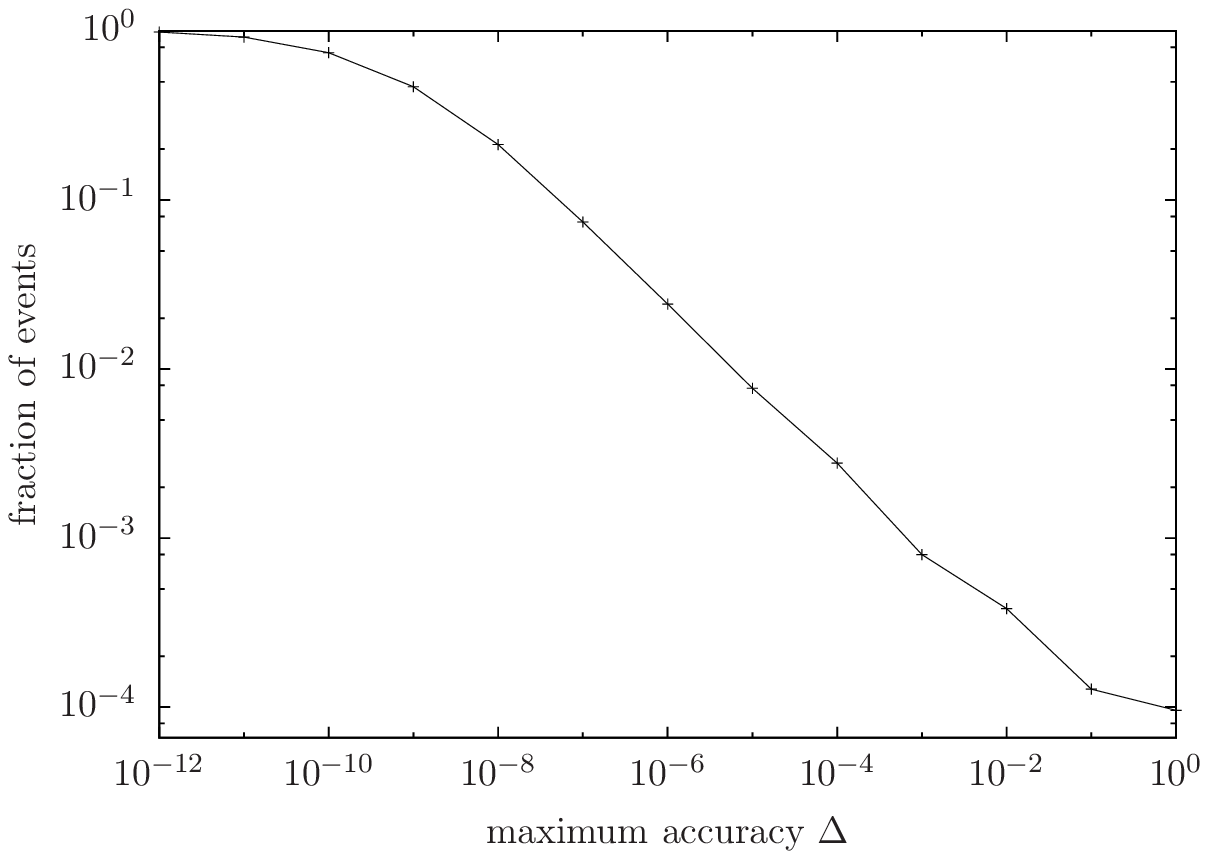}
                \label{plot:gg_ward_identity_check}
        \end{subfigure}
        \vspace*{-3ex}
        \caption[Checks of the virtual contributions]{%
                Checks of the virtual contributions for fixed scale 
                $\mu_0=\mu_\text{fix}$: \subref{plot:mg5_rcl_nlo_agreement} agreement between 
                \recola and \madgraph for the partonic subprocesses $\bar{\Pu}\Pu \to 
                \Pe^+\nu_\Pe \mu^- \bar{\nu}_\mu \Pb \bar{\Pb} \PH$ and $\Pg\Pg\to\Pe^+\nu_\Pe
                \mu^- \bar{\nu}_\mu \Pb \bar{\Pb} \PH$ (left), 
                \subref{plot:gg_ward_identity_check} 
                accuracy in satisfying the Ward identity for $\Pg\Pg\to\Pe^+\nu_\Pe 
                \mu^- \bar{\nu}_\mu \Pb \bar{\Pb} \PH$ (right), both showing the cumulative 
                fraction of events with a relative difference/accuracy larger than $\Delta$.}
\end{figure}

We have performed several comparisons and consistency checks of our
calculation.  We have reproduced the LO hadronic cross sections with
\madgraph~\cite{Alwall:2014hca} using the fixed scale choice. We also
compared the virtual NLO contribution to the squared amplitude,
$2\Re\mathcal{M}^*_0\mathcal{M}_1$, for the $\Pg\Pg$ and $\bar{\Pu}\Pu$
subprocesses computed by \recola~\cite{Actis:2012qn} for many
phase-space points with \madgraph. In
\reffi{plot:mg5_rcl_nlo_agreement} we plot the cumulative fraction of
events with a relative difference larger than $\Delta$ between
$2\Re\mathcal{M}^*_0\mathcal{M}_1$  obtained by
\recola and \madgraph. The median of agreement is about $10^{-7}$ for
the $\Pg\Pg$ subprocess and roughly $5\times10^{-9}$ for the
$\bar{\Pu}\Pu$ subprocess. The agreement is worse than $10^{-3}$ for
less than 0.3\,\% of $\Pg\Pg$- and less than 0.04\,\% of
$\bar{\Pu}\Pu$-subprocess events. For the $\Pg\Pg$ process we checked
the accuracy of virtual matrix elements in satisfying the Ward
identity, by replacing the polarization vector of one of the
initial-state gluons in the one-loop amplitude $\mathcal{M}_1$ by its momentum
normalized to its energy: $\epsilon^\mu_\Pg \to p^\mu_\Pg/p^0_\Pg$. In
\reffi{plot:gg_ward_identity_check} we plot the cumulative
fraction of events with 
$2\Re\mathcal{M}^*_0(\epsilon_\Pg)\mathcal{M}_1(\epsilon_\Pg\to p_\Pg/p^0_\Pg)/2\Re\mathcal{M}^*_0(\epsilon_\Pg)\mathcal{M}_1(\epsilon_\Pg)>\Delta$
for virtual events obtained by \recola with a median
of about $10^{-9}$.

With the Monte Carlo code we developed for the calculation of the
process $\fullProcess$, which employs \recola for the evaluation of
matrix elements, we reproduced the results of \citere{Denner:2012yc}
for the closely related process $\Pp\Pp\to\Pe^+\nu_\Pe
\mu^-\bar{\nu}_\mu\Pb\bar{\Pb}$ for the LHC at $\sqrt{s}=8\TeV$ using
a dynamical scale. Since the Catani--Seymour dipoles
\cite{Catani:1996vz,Catani:2002hc} are the same for the process
without the Higgs~boson, this comparison allows us to test the
implementation of the subtraction formalism. As we use the same
renormalization procedure as in \citere{Denner:2012yc}, this is also
verified via this comparison.  In \citere{Denner:2012yc} the one-loop
matrix elements and the $I$-operator are computed in double-pole
approximation for the two \PW-boson resonances using physical (i.e.\ 
real) \PW- and \PZ-boson masses and $\Gamma_\PW=\Gamma_\PZ=0$.
Moreover, the $M_\PH\to\infty$ limit is adopted, i.e.\ closed fermion
loops involving top quarks coupled to Higgs bosons are neglected. In
contrast, the new code computes the full one-loop matrix elements and
the $I$-operator without any approximation. The largest difference
between both calculations results from the use of the double-pole
approximation for the virtual corrections and is of order
$\alphas\Gamma_\PW/(\pi M_\PW)\sim 0.1\,\%$. We could reproduce the
results of \citere{Denner:2012yc} for the integrated cross section at
NLO within $(2{-}3)\sigma$, which is at the level of 0.5\,\%, thus
confirming both calculations basically within the accuracy of the
numerical integration.

We compared our predictions for the NLO total hadronic cross section
with computations of $\Pt\bar{\Pt}\PH$ production without decays of
the top quarks at NLO for fixed \cite{Beenakker:2002nc} and dynamical
scale \cite{Frederix:2011zi}. To this end we performed NLO
computations using the references' parameters and PDF mappings and a
minimal set of cuts on the decay products of the top and antitop in
our process.  Since our calculation includes background processes to
$\Pt\bar{\Pt}\PH$ production, a minimal set of cuts has to be
maintained to ensure infrared safety: to avoid large contributions
from possible collinear events with bottom quarks in forward direction
(as induced by diagrams shown in
\reffi{fig:born_notops_gg_tchannel_z0} or
\reffi{fig:born_notops_gg_tchannel}) we kept a small \Pb-jet
transverse-momentum cut ($\ptsub{\Pb}>2\GeV$) and a small \Pb-jet
distance cut ($\Delta R_{\Pb\Pb}> 0.01$) to avoid singular events from
gluon splitting ($\Pg\to \Pb\bar{\Pb}$) as in
\reffi{fig:born_notops_uxu_tchannel}. We multiply appropriate branching ratios to
the results of \citeres{Beenakker:2002nc,Frederix:2011zi} in the
narrow-top-width approximation (NtWA), i.e.\ 
$\int\text{d}\sigma_\text{NtWA}=\sigma_{\Pt\bar{\Pt}\PH}\text{BR}_{\Pt\to
  i}\text{BR}_{\bar{\Pt}\to j}$, and apply the NLO matching prescription of
Section 2.1.2 in \citere{Denner:2012yc}  to our results.  Thus,
we find agreement of our NLO predictions with
\citere{Beenakker:2002nc} and \citere{Frederix:2011zi} within 1\,\%,
which is of the expected order of $\Gamma_\Pt/m_\Pt$, since our
calculation includes off-shell and non-resonant top-quark effects.

\section{Conclusions}
\label{sec:Conclusions}

In this article we have presented the calculation of the
next-to-leading-order QCD corrections to off-shell top--antitop
production in association with a Higgs boson with leptonic decay of
the top quarks at the LHC, including all resonant, non-resonant, and
off-shell effects of top quarks as well as all interferences. For the
computation of leading- and next-to-leading-order matrix elements we
have utilised the recursive matrix-element generator \recola linked to
the one-loop integral library \collier. The phase-space integration
has been performed with an in-house multi-channel Monte-Carlo code
that implements the dipole subtraction formalism.

We provided integrated cross sections and several differential
distributions for a 13\TeV LHC using a fixed and a dynamical scale
that have both been used in the literature for the computation of NLO
QCD corrections of $\Pt\bar\Pt\PH$ production with a stable Higgs
boson and stable top quarks. We find almost the same integrated cross
sections and scale dependence for both scale choices at leading order
as well as next-to-leading order QCD, with \change{a similar $K$~factor of
1.172 and 1.176 for the dynamical and fixed scale choice, resp.} However, 
the use of the dynamical scale instead of the fixed
scale improves the perturbative stability in high-energy tails of most
distributions, especially those of transverse momenta. Using the
dynamical scale, we find $K$~factors in the range $1.0{-}1.4$ and
residual scale uncertainties are the level or $10\,\%$ for
distributions.  For the integrated cross section an extrapolation to
the zero-top-width limit has been performed, indicating
 non-resonant and off-shell top-quark effects below one per cent.
\change{While this effect is small, our calculation is also the first
  one to include NLO correction effects in the
  top--antitop--Higgs-boson production and the top decay processes.}

Besides its phenomenological relevance, this calculation demonstrates
the power of the tools \recola\ and \collier\ in performing
complicated NLO calculations.

\acknowledgments 
This work was supported by the Bundesministerium
f\"ur Bildung und Forschung (BMBF) under contract no. 05H12WWE.

\bibliographystyle{JHEP}
\bibliography{ttxh_nlo} 

\end{document}